\documentclass[a4paper,preprintnumbers,floatfix,superscriptaddress,pra,twocolumn,showpacs,notitlepage,longbibliography]{revtex4-1}
\usepackage[utf8]{inputenc}
\usepackage[T1]{fontenc}
\usepackage[sc,osf]{mathpazo}
\usepackage{amsmath}
\usepackage{amsthm}
\usepackage{bbm}
\usepackage{comment} 
\usepackage{marvosym}
\usepackage{graphicx}
\usepackage{xcolor}
\usepackage{appendix}
\usepackage{hyperref}

\usepackage{float}
\newcommand{\ket}[1]{| #1 \rangle}
\newcommand{\bra}[1]{\langle #1 |}

\newcommand{\beq}{\begin{eqnarray}}
\newcommand{\eeq}{\end{eqnarray}}

\newcommand{\mean}[1]{\ensuremath{\langle{#1}\rangle}}

\newtheorem{theorem}{Theorem}

\newtheorem{result}[theorem]{Result}

\DeclareMathOperator{\tr}{\mathrm{tr}}

\def\1{\mathbf{1}}
\def\0{\mathbf{0}}
\newtheorem{observation }{Observation}
\begin{document}

\title{High Schmidt number concentration in quantum bound entangled states}

\author{Robin Krebs}
  \email{robinbenedikt.krebs@stud.tu-darmstadt.de}
 \affiliation{
 Department of Computer Science, Technical University of Darmstadt, Germany
}
\author{Mariami Gachechiladze}%
 \email{mariami.gachechiladze@tu-darmstadt.de}
\affiliation{
 Department of Computer Science, Technical University of Darmstadt, Germany
}

\begin{abstract}
A deep understanding of quantum entanglement is vital for advancing quantum technologies. The strength of entanglement can be quantified by counting the degrees of freedom that are entangled, which results in a quantity called Schmidt number. A particular challenge is to identify the strength of entanglement in quantum states which remain positive under partial transpose (PPT), otherwise recognized as undistillable states. Finding PPT states with high Schmidt number has become a mathematical and computational challenge. In this work, we introduce efficient analytical tools for calculating the Schmidt number for a class of  bipartite states, called generalized grid states. Our methods improve the best known bounds for PPT states with high Schmidt number. Most notably, we construct a Schmidt number three PPT state in five  dimensional systems and a family of states with a Schmidt number of $(d+1)/2$ for odd $d$-dimensional systems, representing the best-known scaling of the Schmidt number in a local dimension. Additionally, these states possess intriguing geometrical properties, which we  utilize to construct indecomposable entanglement witnesses.
\end{abstract}

\maketitle

\textit{Introduction} -- Quantum entanglement is a fundamental phenomenon of quantum theory on which success of the rapidly advancing field of quantum technologies relies. However, in spite of the development of a complex mathematical theory of entanglement, with the primary goal to detect and quantify the entanglement present in a physical system, many fundamental questions still need to be answered. So far, no efficiently computable necessary and sufficient criterion for separability of a composite quantum state has been discovered. Generally, the problem is known to be NP-hard~\cite{gurvits2003classical, gharibian2015quantum, aliferis5authors}, and it has only been solved in qubit-qubit and qubit-qutrit cases using the famous positive partial transpose criterion~\cite{peres1996separability,HORODECKI19961}. In higher-dimensional states, however, there is no general method to investigate entanglement.

At the same time,  in recent years, high-dimensional entanglement has become experimentally feasible~\cite{dada2011experimental,malik2016multi}, demonstrating a better noise resistance in a number of applications, in comparison with low-dimensional implementations~\cite{huber2013weak,mirhosseini2015high,wang2005quantum,lanyon2009simplifying}. As a result, determining whether an experiment successfully established high-dimensional entanglement or if the experimental results can be explained by assuming low-dimensional entanglement is critical. A go-to measure which certifies that a bipartite state has been entangled across at least $r$ degrees of freedom is called \textit{Schmidt number}~\cite{terhal2000schmidt}, and is very challenging to estimate due to its form (see below) for general states. One typical way to determine this quantity is to construct a hermitian operator, called a \textit{Schmidt number witness}, a few examples of which can be found in literature~\cite{sanpera2001schmidt,wyderka2023construction,bavaresco2018measurements}.

An additional challenge lies into constructing a Schmidt number witness for states with positive partial transpose (PPT states), since such hermitian operators have to satisfy an extra mathematical property, known as being \textit{indecomposable}~\cite{chruscinski2008construct}. Moreover, such states are known to be bound entangled, as no pure state entanglement can be ever distilled from them~\cite{horodecki1999bound}.  Because of this property, it was originally believed that PPT states are weakly entangled and cannot be used for quantum information processing tasks~\cite{peres1999all}. Yet, contrary to this perception, a family of PPT states with logarithmically increasing Schmidt number in the local dimension were found~\cite{chen2017schmidt}. Later, along with discovering the potentialities of bound entangled states in quantum steering, nonlocality, and secure communication~\cite{moroder2014steering,vertesi2014disproving, horodecki2005secure}, PPT states with the Schmidt number scaling of $\frac{d}{4}$ and $\frac{d}{2}$ were proposed in even dimensional $d\times d$ systems~\cite{huber2018high,pal2019class}. At the same time, upper bounds have been derived on the amount of entanglement in bound entangled states. For example, in $3\times3$ no PPT state exists with Schmidt number three~\cite{yang2016all}. In sequences of works, high Schmidt number PPT states have been extensively investigated~\cite{huber2018high,pal2019class, sanpera2001schmidt,cariello2017gap,cariello2020inequalities,marciniak2021class} and searched for. Despite these developments in the study of bound entanglement, systematic tools to construct Schmidt number entanglement witnesses for PPT states are majorly absent. 

In this work, we address precisely this problem and for an elegant class of quantum states, so called \textit{generalized grid states}~\cite{lockhart2018entanglement, ghimire2023quantum},  we develop a set of efficient graphical tools relying on a generalized range criterion to evaluate their exact Schmidt number. Here, we provide examples of PPT states that enjoy the highest known Schmidt number \textit{concentration} in given local dimensions. To start with, despite the efforts, the best minimal example of a PPT bound entangled state with Schmidt number three was known to be in local dimensions $6\times 6$~\cite{pal2019class}. Here we find a PPT state with Schmidt number three in $5\times 5$ systems. We also improve other previously known bounds and derive a family of PPT states with Schmidt number scaling $(d_A+1)/2$ in $d_\mathrm{A}\times d_\mathrm{B}$ dimensional systems, where  $d_\mathrm{A}$ is odd and $d_\mathrm{A}<d_\mathrm{B}$. We find these states by resorting to tools similar to entanglement distillation, but in this  case, with the goal to distil a high Schmidt number PPT state from multiple copies of low Schmidt number ones. We call this procedure Schmidt number \textit{concentration}. 
 Similar tasks have been studied for quantum Fisher information~\cite{pal2021bound} and nonlocality~\cite{tendick2020activation}. 
As a last example, using the Schmidt number concentration, we find a Schmidt number three state in $4\times 12$ systems. Our findings leave as an open problem whether there exist Schmidt number three states 
in $3\times d$, and what the smallest local dimension is in $4\times d$  systems to accommodate such states.

Finally, our new states are not only highly entangled, but they all enjoy a striking property of being extremal in a PPT set, that is, they cannot be decomposed as a convex mixture of other PPT states. We utilize this property to obtain indecomposable entanglement witnesses for high Schmidt number states and numerically obtain such witnesses for all such states. While our findings are specifically tailored to generalized grid states, the methods developed in this study can be extended to a broader range of quantum systems. This paves the way for exploring new signatures of high dimensional entanglement and bound entanglement.

\emph{Preliminaries} -- Given a bipartite quantum state $\rho_{AB}$ defined over a finite-dimensional Hilbert spaces $\mathcal{H}_A\otimes \mathcal{H}_B$,  and the transposition map $T$ on one of the subsystems,  we say that the state is positive under partial transposition or is PPT, if  
\begin{equation}
    \rho_{AB}^{T_B}\geq 0,\  \text{ where  }\  \rho_{AB}^{T_B}:= \mathbbm{1}_A \otimes T_B (\rho_{AB}).
\end{equation}
Any state that is negative under the partial transposition map is entangled and, thus, can be detected via a hermitian operator $W$ called an entanglement witness~\cite{guhne2009entanglement}, and moreover, by a so-called \textit{decomposable} entanglement witness of the form $W=P+Q^{T_\mathrm{B}}$, with some $P,Q\ge 0$. 
However, such decomposable witnesses do not detect  PPT entanglement, instead one needs to construct an \textit{indecomposable} witness.  The Schmidt number (SN) of a state $\rho_{AB}$ is equal to the smallest $k$, such that there exists a decomposition of $\rho_{AB}$ into a set of pure states of Schmidt Rank (SR) at most $k$ (SR corresponds to a number of nonzero Schmidt coefficients)~\cite{terhal2000schmidt}.
Several SN witnesses have been derived in literature~\cite{sanpera2001schmidt,wyderka2023construction,bavaresco2018measurements}, but determining the SN remains a difficult task, in particular for PPT states, for which  indecomposable SN witnesses have to be constructed.

A well-known method to determine the SN is through a generalized range criterion~\cite{sanpera2001schmidt}. The range of a density matrix $\rho_{AB}$ is defined as the image of $\rho_{AB}$,  $R(\rho_{AB}): = \{\rho\ket{\psi}\ |\ \ket{\psi} \in \mathcal{H}_A\otimes\mathcal{H}_B\}$. 
Next, we define the Schmidt rank restricted range of a density matrix $R_k(\rho_{AB}):=\{\mathrm{SR}(\ket{\psi})\le k\ |\ \ket{\psi}\in R(\rho_{AB})\}$, and give a generalized range criterion for SN:
All SN $k$ states must have a complete basis in $k$-restricted range $R_k(\rho_{AB})$ to span $R(\rho_{AB})$. Then, if there is a vector $\ket{v}\in R(\rho_{AB})$ which is orthogonal to its $k$-restricted range, $\ket{v}\perp R_k(\rho_{AB})$, the state $\rho_{AB}$ must have a Schmidt number greater than $k$.

\begin{figure}
    \centering
\includegraphics[width=0.45\textwidth]{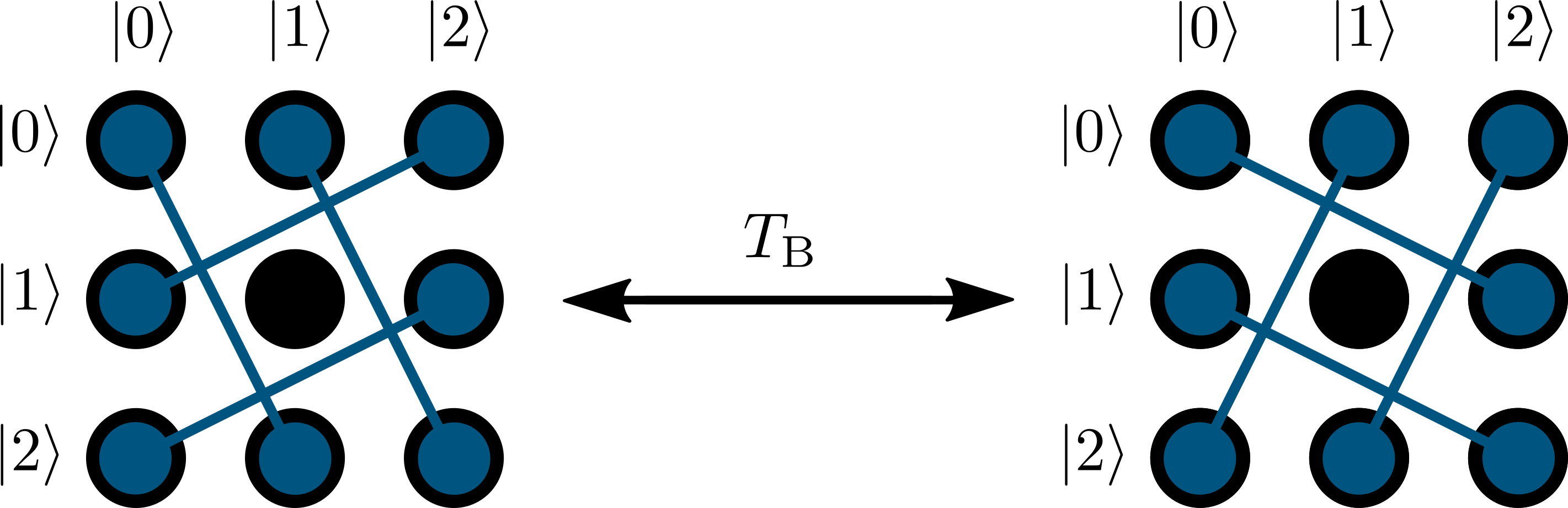}
    \caption{(Left) 
    The crosshatch state, $\rho^{\mathrm{CH}}$~\cite{lockhart2018entanglement}, a bound entangled grid state in $3\times 3$. The vertices in the grid are labeled as $\{(0,0),(0,1),(0,2),(1,0),(1,1),(1,2),(2,0),(2,1),(2,2)\}.$ The crosshatch state has edges $E=\{\{(0,0),(2,1)\},\{(0,1),(2,2)\}$, $\{(1,0),(0,2)\},\{(2,0),(1,2)\}\}.$  (Right) Partial transpose of the crosshatch state, which is also a valid grid state.} 
    \label{fig:crosshatch}
\end{figure}

\begin{figure}[t]
    \centering    \includegraphics[width=0.47\textwidth]{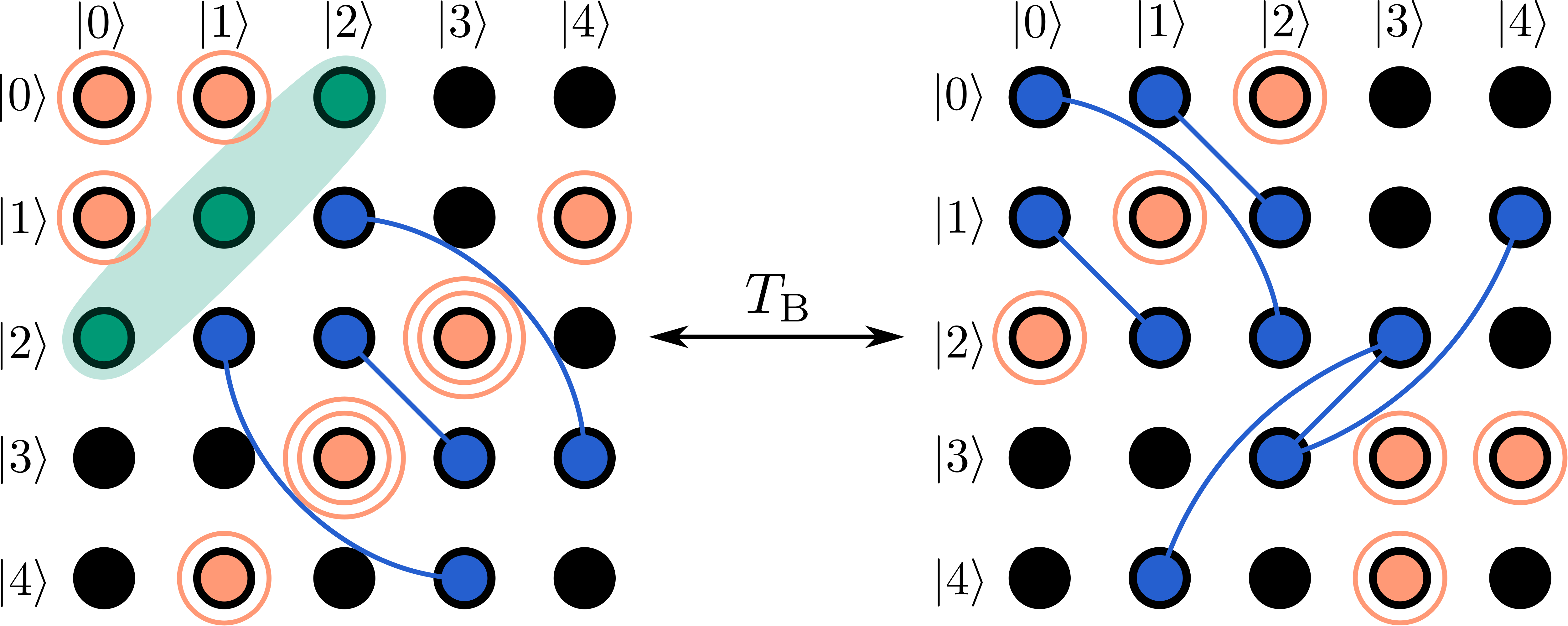}
    \caption{(Left) The smallest known Schmidt number three bound entangled state in $5\times 5$, $\rho^{5,5}$, with the multi-set of edges $E=\{\{(0,0)\},$ $\{(0,1)\},$ $\{(1,0)\},$ $\{(2,3)\},$ $\{(2,3)\},$ $\{(3,2)\},$ $\{(3,2)\},$ $ \{(1,4)\},$ $\{(4,1)\},$ $\{(2,1), (4,3)\},$ $
    \{(2,2),(3,3)\},$ $
    \{(1,2),(3,4)\},$ $ 
    \{(0,2),(1,1),(2,0)\}\}$.
    (Right) The grid state corresponding to a partial transposition of $\rho^{5,5}$ has Schmidt number two. 
    }
    \label{fig:5dbe}
\end{figure}

\textit{Generalized Grid States and SN criterion} -- Generalized grid states are mixed quantum states with elegant graphical representation (see Figs.~\ref{fig:crosshatch}, \ref{fig:5dbe} for examples).  
Following the original definition~\cite{lockhart2018entanglement,ghimire2023quantum}, we first define a $d_A\times d_B$ grid, $G_{\mathrm{AB}} = \{(i,j)\ |\ 0\le i < d_A,0\le j < d_B\}$, considered as a set of vertices, with a computational basis state of the bipartite $d_A \times d_B$ Hilbert space $\mathcal{H}_{\mathrm{A}}\otimes \mathcal{H}_{\mathrm{B}}$ assigned to every vertex $(i,j)\mapsto \ket{ij}$. A generalized grid state then corresponds to a \textit{non-simple } hypergraph $H=(V,E)$, where the set of vertices is $V\subseteq G_{\mathrm{AB}}$, and every edge $e$ in the multiset $E$ contains a subset of vertices and corresponds to the unnormalized superposition $\ket{e} = \sum_{v\in e}\ket{v}$, called edge representation. If an edge contains $k$ vertices, we call it a $k$-edge, $1$-edges are called loops.   A generalized grid state $\rho_H$ corresponding to a hypergraph $H$ is then defined as the equal mixture of its edge representations, 
\begin{equation}
    \rho_H = \frac{1}{\tr{\big(\sum_{e\in E} \ket{e}\bra{e}\big)}}\sum_{e\in E} \ket{e}\bra{e}.
\end{equation}
Finally, a vertex $v\in G_{\mathrm{AB}}$ is called \textit{isolated}, if $v\notin V$. It corresponds to a product state in the kernel of $\rho_H$, playing an important role later.

One of the main advantages of defining grid states with $k$-edges, $k>2$, is the ability to construct high Schmidt number quantum states. 
In this work, we focus on generalized PPT grid states whose partial transposition also is a grid state. 
In general, deciding if a given density matrix is a grid state is a hard problem ~\cite{letchford2012binary}.
Here we give a simple sufficient criterion which we know is not necessary but is powerful enough to find interesting PPT states. Given a generalized grid state $\rho_H$, to decide if it is PPT, split it into its diagonal part $D$, and off-diagonal part $A$, regarded as an adjacency matrix of the graph $G_H$, constructed by replacing every edge $e\in E$ in hypergraph $H$ by a complete graph on all vertices in $e$. 
The action of the partial transposition on $\rho_H$ corresponds to flipping every edge $\{(i_1, j_1),(i_2, j_2)\}$  in the graph $G_H$ to $\{(i_1, j_2),(i_2, j_1)\}$. 
The graph formed by these flipped edges is denoted as $G^{T_B}_H$.
Then, given $2$-colorability of $G^{T_B}_H$, $\rho_{H}^{T_\mathrm{B}}\ge 0$ iff the degree of every vertex in $G^{T_\mathrm{B}}_H$ is not higher than the correcponding diagonal entry in $D$. See Figs.~\ref{fig:crosshatch} and~\ref{fig:5dbe} (left)  for examples and Appendix~\ref{app:PPT_Crit} for the proof. 

Next, we show how to calculate the Schmidt number using the generalized range criterion. For a generalized grid state $\rho_H$, we have that
any $\ket{\psi}\in R(\rho_H)$ can be expressed as an arbitrary linear combination of 
edge representations, $\ket{\psi} = \sum_{e\in E} c_e\ket{e}$. 
Furthermore, $\mathrm{SR}(\ket{\psi})\le k$ if and only if
the corresponding coefficient matrix $\Psi$ has all of its size $(k+1)$ minors equal to zero, where $\{\Psi\}_{ij}:=\mean{ij|\psi}$, and $\{\ket{i}\}_{i=0}^{d_A-1}$ and $\{\ket{j}\}_{j=0}^{d_B-1}$ are the computational bases~\cite{cubitt2008dimension,chen2006quantum}.

As a simple example, we apply this criterion to find the SN of the $3\times 3$ crosshatch state $\rho^\mathrm{CH}$  in Fig.~\ref{fig:crosshatch} and defined originally in Ref.~\cite{lockhart2018entanglement}. First, we give an explicit parametrization of the range of $\rho^\mathrm{CH}$, 

\begin{equation}
\label{eq:ch_range}
\Psi^\mathrm{CH}= 
    \begin{pmatrix}
        c_{00} & c_{01} & c_{10}    \\
        c_{10} & 0 & c_{20}  \\
        c_{20} & c_{00} & c_{01} 
    \end{pmatrix} \in R(\rho^{\mathrm{CH}}), 
\end{equation}
and show that there is no product vectors contained in its range.  To that end, we first determine the $1$-restricted range $R_1(\rho^{\mathrm{CH}})$ by demanding that all the $2\times2$ minors of $\Psi^{\mathrm{CH}}$ vanish. Four conditions of the $2\times2$ vanishing minors have a particularly simple form: 
\begin{equation} \label{eq:3by3vanishing_minor}
    c_{01}c_{10}= c_{01}c_{20}= c_{00}c_{10} = c_{00}c_{20} =0,
\end{equation}
admitting two nontrivial solutions, either $c_{10}=0$ and $c_{20}=0$ or $c_{00}=0$ and $c_{01}=0$. These cases are equivalent under relabeling, and assuming $c_{10}=c_{20}=0$, we obtain another set of $2\times 2$ minor equations $c_{00}^2 = c_{01}^2 = 0$,
implying that $c_{00},c_{01}=0$, and hence that $\Psi^\mathrm{CH}=0_{3\times 3}$, and thus there is no nonzero vector in the $1$-restricted range of  $\rho^\mathrm{CH}$, proving that the state must have $\mathrm{SN}(\rho^\mathrm{CH})=2$.

The simple form of the range criterion, we used here, entirely relies on having isolated vertices in a state, as this construction  translates  
to the powerful vanishing minor conditions in Eq.~(\ref{eq:3by3vanishing_minor}). In the same way, when testing a generalized range criterion for $k$-restricted range, if we want to keep the conditions simple, we consider states which result in the $(k+1)\times (k+1)$ minor vanishing conditions having the following form, $\prod_{i=0}^{k} c_{e_i} = 0$ for $(k+1)$ distinct coefficients. 
In this case, one of the $(k+1)$ coefficients shall be zero, leading to  possible solution branches. We represent each branch as a generalized grid state, with the respective edge $e_i$ erased from the grid state. This procedure is repeated on all the branches and their subbranches, until no more edges can be erased in this manner. If an edge $e\in E$, with $\mathrm{SR}(\ket{e})=(k+1)$, 
can be erased in all solution branches, then $\ket{e}\perp R_k(\rho_H)$, implying $\mathrm{SN}(\rho_H)>k$. 

We are ready to report on the first result using the construction described above and the generalized range criterion.

\begin{result}  A Schmidt number three PPT bound entangled state exists in the local dimension $5\times 5$.
\end{result}

The smallest Schmidt number three state was believed to be in the local dimension $6\times 6$~\cite{pal2019class}. See Fig.~\ref{fig:5dbe} for the example of the Schmidt number three PPT state we found using our construction, $\rho^{5,5}$ in  $5\times 5$ and Appendix \ref{app:55app} for the proof.

\begin{figure}
    \centering
\includegraphics[width=0.49\textwidth]{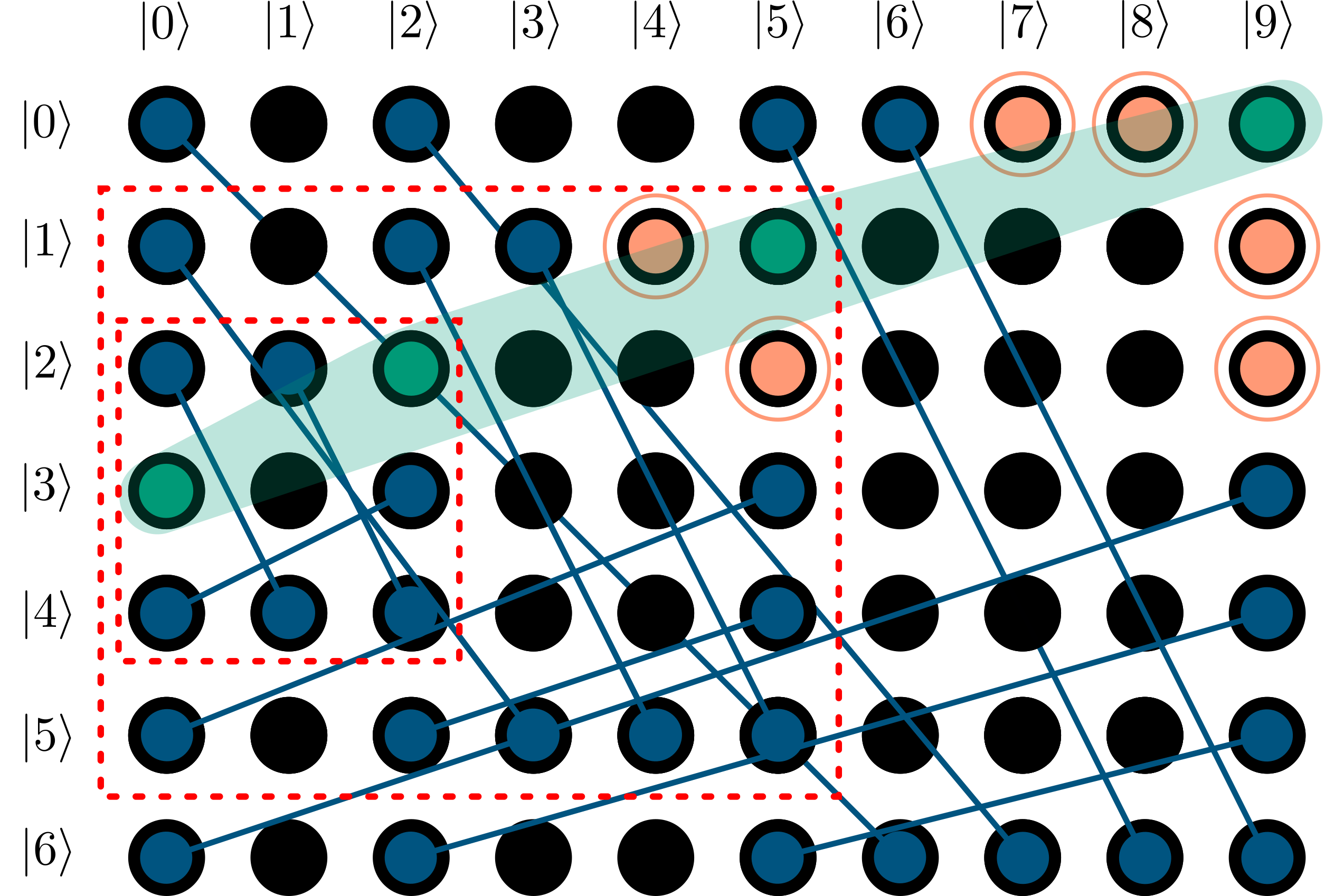}
    \caption{
    Examples of $\rho^{(n)}$, for $n<4$. Observe the nested structure of the examples, as indicated by the dotted rectangles. The unique Schmidt rank $>2$ component is indicated as shaded shape, Schmidt rank 2 as lines, and separable as circles. The innermost rectangle contains the crosshatch state, the next one contains the first example of a SN three state in local dimension $5\times 6$, while the whole figure shows SN four in $7\times 10$.}
    \label{fig:snfamily}
\end{figure}

\emph{High Schmidt number concentration} -- It is well known that no entanglement can be distilled from the PPT entangled states. Instead, here we use the tools from a distillation protocol to achieve \textit{Schmidt number concentration} in bound entanglement. The idea is to use different low Schmidt number PPT states (relative to the local dimension), act on them with local filtering operations,  and obtain a single PPT state with high relative Schmidt number for the resulting local dimension. Clearly, it is important to tailor the concentration protocol to particular PPT states to make sure efficiency of calculating SN in the resulting distilled state. In what follows, we describe the procedure.

Let $\rho_{A_1B_1}=p\sigma+(1-p)\ket{v}\bra{v},$ be a bipartite Schmidt number $k$ state, $0<p<1$, and let $\ket{\alpha\beta}$ be a product state and $\ket{v}$ a pure state fulfilling the following technical conditions:
$\ket{v},\ket{\alpha\beta}\perp R(\sigma),~\ket{v}\perp R_{k-1}(\rho),~ \ket{v}\not\perp \ket{\alpha\beta}$, then there exists a PPT preserving map $\Theta$ with its action on a bipartite state $\rho_{A_1B_1}$ defined as follows, 
\begin{align}
    \Theta(\rho_{A_1B_1}) =&\left(A\otimes B\right)\rho_{A_1B_1}\otimes\rho^{\mathrm{CH}}_{A_2B_2}\left(A^\dagger\otimes B^\dagger\right),
\end{align}
where $A$ and $B$ are local filtering operators acting on the corresponding local Hilbert spaces. The resulting state is a bound entangled state, $\tilde\rho_{AB}$ with the same properties as $\rho_{AB}$ but with the Schmidt number increased by one, $\mathrm{SN}(\tilde\rho_{AB})=(k+1)$, and
\begin{align}
  \tilde\rho_{AB} = \tilde p \tilde\sigma+(1-\tilde p)\ket{\tilde v}\bra{\tilde v}.
\end{align}
In Appendix~\ref{app:snfamily}, we 
define operations $A,~B$ and
prove that if the original state $\rho_\mathrm{AB}$ has local dimension $d_\mathrm{A}\times d_\mathrm{B}$, and Schmidt number $k$, the map $\Theta$ changes those quantities to: 
\begin{equation}
    (d_\mathrm{A}, d_\mathrm{B},k)\mapsto (d_\mathrm{A}+2, \ d'_\mathrm{B},\ k+1),
\end{equation}
where $d'_\mathrm{B} = d_\mathrm{B}+\mathrm{rank}\left(\left(\bra{\alpha}_\mathrm{A}\otimes\mathbb{1}_\mathrm{B}\right)\rho_\mathrm{AB}\left(\ket{\alpha}_\mathrm{A}\otimes\mathbb{1}_\mathrm{B}\right)\right)+1$.
Since the resulting state shares the same properties as the starting one, we can repeatedly apply the map $\Theta^n(\rho_{A_1,B_1})$, and study the Schmidt number concentration for increasing $n$. The exact values for scaled local dimensions and Schmidt number are derived in Appendix~\ref{app:distil_Grid}. 
If we apply the concentration protocol to the crosshatch state $\rho^\mathrm{CH}$, we transform $n$ copies of $3\times 3$ crosshatch states each with Schmidt number $2$ into a bound entangled generalized grid state $\rho^{(n)}=\Theta^{(n-1)}(\rho^\mathrm{CH}_{A_1B_1})$ with properties the $(2n+1,~(n+1)(n+2)/2,~ n+1)$, giving our second result.

\begin{result}~\label{res:concentration}
    For odd values of $d$, Schmidt number $\frac{(d+1)}{2}$ PPT states exist in the local dimension $d\times \frac{(d+1)(d+3)}{8}$.
\end{result}
The Schmidt number scaling present in Result~\ref{res:concentration} improves on the best known construction in even-dimensional systems with the Schmidt number scaling as $\frac{d}{2}$~\cite{pal2019class}.  
See Appendix~\ref{app:distil_Grid} for the construction and proof of the protocol, and Fig.~\ref{fig:snfamily} for $\rho^{(n)}$, $n\leq 3$.

Our Schmidt number concentration technique is not limited to grid states only. To show this, we design a SN concentration protocol for a Horodecki state $\rho^{\mathrm{Hor}}$ in $2\times 4$, the lowest dimensional bound entangled state in nonhomogeneous local dimensions~\cite{horodecki1997separability}, and obtain a Schmidt number three bound entangled state in $4\times 12$,
\begin{equation}
    \rho^{4,12} =  \left(\Pi_{\mathrm{A}}\otimes\mathbb{1}_{\mathrm{B}}\right)\rho^{\mathrm{Hor}}_{\mathrm{A}_1\mathrm{B}_1}\otimes\rho^{\mathrm{CH}}_{\mathrm{A}_2\mathrm{B}_2}  
    \left(\Pi_{\mathrm{A}}\otimes\mathbb{1}_{\mathrm{B}}\right)^\dagger, 
\end{equation}
where $\Pi_{\mathrm{A}} := \mathbb{1}_{\mathrm{A}_1\mathrm{A}_2}-\ket{1}\bra{1}_{\mathrm{A}_1}\otimes\left(\ket{0}\bra{0}+\ket{2}\bra{2}\right)_{\mathrm{A}_2}$.
See Appendix~\ref{app:4x12} for proof techniques and the explicit form of the states $\rho^{\mathrm{Hor}}$ and $\rho^{4,12}$.

\textit{Extremal PPT states and indecomposable witnesses} -- We call  a PPT state $\rho$ \textit{extremal} if it cannot be decomposed into a convex mixture of a pair of distinct PPT states. 
This means that subtracting any PPT state with a Schmidt number less than $\mathrm{SN}(\rho)$ results in a matrix with negative partial transpose, a property that we use to construct SN witnesses for such states, specifically making use of the ranges $R(\rho)$, $R(\rho^{T_\mathrm{B}})$.
It is worth mentioning at this point, that all PPT entangled extremal states belong to
the intersection of faces of the positive semidefinite set and its partial transpose, but are not extremal states of both intersected sets.
Moreover, such a PPT entangled extremal state is the unique state supported on the ranges $R(\rho)$ and $R(\rho^{T_\mathrm{B}})$~\cite{leinaas2007extreme}. We can directly use this result to obtain Schmidt number PPT entanglement witnesses:  
 
Take the projectors $P$ and $Q$ associated to the subspaces $R(\rho)$ and $R(\rho^{T_\mathrm{B}})$, and add them in the following manner $W_\rho := P+Q^{\mathrm{T_\mathrm{B}}}$.
Then $\tr(W_\rho \sigma )\le 2$, for all PPT $
\sigma$, with equality holding for $\sigma=\rho$ only. 
Then for every PPT extremal $\rho$ with $\mathrm{SN}(\rho)=k+1$, we can find $\mu_k$ such that $\mu_k:=\max_{\sigma_k}\tr(W_\rho \sigma_k )<2$   
for  PPT $\sigma_k$ with the Schmidt number $k$.  
Then $W_{\mathrm{ind},\rho} := (\mu_k\mathbb{1}-W_\rho)$ is a valid witness for Schmidt number $(k+1)$ PPT states on the PPT set. It can be easily confirmed that $W_{\mathrm{ind},\rho}$ is indeed an indecomposable witness as long as $\mu_k<2$. More details about this construction can be found in Appendix~\ref{app:snedgestates}.

We can use this technique to construct indecomposable entanglement witnesses on the PPT set, as all new states $\rho_i$ depicted 
in this work (in Figs. \ref{fig:crosshatch}, \ref{fig:5dbe}, \ref{fig:snfamily}, and \ref{app:fig:4x12sn3}) are extremal  PPT states. 
For those states, we determine lower  on the corresponding  $\mu^i_1$ using local optimization \cite{ling2010biquadratic} and  upper bounds utilizing the DPS hierarchy \cite{doherty2004complete}. We report our numerical results and confidence intervals in Table~\ref{tab:mutable}. 
See Appendices~\ref{app:extremetest},\ref{app:numrange} for further details.  
\begin{table}
    \begin{tabular}{c|cccccc}
         & $\rho^{\mathrm{CH}}$
         &$\rho^{5,5}$ 
         & $\rho^{(2)}$ & $\rho^{(3)}$ & $\rho^{4,12}$\\
         \hline
         $\mu_\mathrm{L}$ & $\approx 1.8659$  &$1.9$ & 
         $\frac{5+\sqrt{5}}{4}$ & $\frac{5+\sqrt{5}}{4}$ & $\frac{5+2\sqrt{2}}{4}$\\
        $\mu_\mathrm{U}-\mu_\mathrm{L}$\  
        & \ $9\times10^{-5}$\
        &  \ $9\times 10^{-4}$\  &  
        \ $4\times10^{-8}$\  & $3.6\times10^{-2}$ & $5\times10^{-10}$ \\
    \end{tabular}
    \caption{Table containing upper and lower bounds of $\mu_1$ for various states. 
    A local minimum was determined from the numerical lower bounds and upper bounds by DPS hierarchy.}
    \label{tab:mutable}
\end{table}

\emph{Discussion} --  Increasing attention has been directed towards high-dimensional entanglement, primarily due to its potential advantages in robust quantum information processing. Even in the bipartite scenario, understanding the nature of such entanglement is still a difficult task beyond the basic dimensions of two-qubit or qubit-qutrit systems.  The particular challenge lies in identifying states that are both highly entangled and positive under partial transposition (PPT). In this study, we explored novel classes of highly entangled bound entangled states, yielding several technical and conceptual advancements. We devise an efficient tool to accurately estimate the Schmidt number for generalized PPT grid states. Employing this tool, we identify states with the highest known Schmidt number within fixed local dimensions. Additionally, by applying established entanglement distillation techniques to PPT states, we achieve high Schmidt number concentration in bound entanglement. Consequently, we established a family of states with the best-known Schmidt number scaling with nonhomogeneous local dimensions. Finally, we demonstrated that our states exhibit the remarkable property of extremality within the PPT set, a feature we leveraged to construct indecomposable Schmidt number witnesses on a PPT set. 

There are a few natural directions to extend our work. First, on the application side, interestingly nearly all our PPT Schmidt number $k$ states enjoy the property that they can be expressed as convex mixture of a Schmidt rank $k$ maximally entangled state and Schmidt number $2$ states supported on the non-overlapping range. This particular property renders our states intriguing contenders as potential counterexamples to the still unresolved PPT$^2$ conjecture. From a technical perspective, the characterization and detailed examination of the generalised grid states could be beneficial since they constitute interesting PPT states. 
Moreover, the family of states presented in this work constitute extremal points of the PPT set. It is interesting to explore geometrical undertones for this connection, and use the states to construct indecomposable witnesses for high Schmidt number detection.  Finally, for the Schmidt number concentration protocol, we developed a graphical language to depict the copies of the mixed states. This approach could help explore properties of quantum states which are not known to be closed under taking the tensor product.

\emph{Acknowledgments}--  We thank Otfried G\"uhne  and Yuanyuan Mao for the suggestions and discussions at the original stage of the project and Armin Tavakoli, Simon Morelli, Gernot Alber, and Nikolai Miklin for discussions.

\onecolumngrid

\section*{Appendix}
The following appendices \ref{app:PPT_Crit}-\ref{app:4x12} contains proofs and technical derivations. Appendix~\ref{app:PPT_Crit} proves a sufficient PPT criterion for the generalized grid states. Appendix \ref{app:55app} contains the proof that $\rho^{5,5}$ has Schmidt number three. Appendix~\ref{app:snfamily} explains how $\Theta$ is constructed, and proves that it can increase the Schmidt number efficiently given a set of conditions, while the next Appendix~\ref{app:distil_Grid} explains how iterated application of $\Theta$ may be used to construct a family of grid states with high Schmidt number scaling (see Fig.~\ref{fig:snfamily}), alongside a quantitative discussion of the exact scaling of Schmidt numbers and dimensions. Appendix~\ref{app:4x12} contains the proof that $\rho^{4,12}$ has Schmidt number 3. In Appendix~\ref{app:snedgestates} the concept of edge states is generalized to high Schmidt numbers and the relationship to Schmidt number witnesses and the range criterion is further discussed. The method used for confirming some examples as extremal states of the PPT set is explained in appendix~\ref{app:extremetest}.
Appendix \ref{app:numrange} explains the methodology for obtaining the numerical results in table \ref{tab:mutable}. 

\begin{appendix}
\section{Sufficient graphical PPT criterion for generalized grid states}\label{app:PPT_Crit}

\begin{proof}
The PPT criterion, stating that a hypergraph $H$ with a $2$-colorable transposed graph $G^{T_\mathrm{B}}$ corresponds to a PPT state iff $G^{T_\mathrm{B}}$ does not have a higher degree at every vertex than $H$, can be proven by observing that a $2$-colorable graph does not contain fully connected subgraphs on more than two vertices. Any fully connected component on three or more vertices contains a triangle, and is therefore not $2$-colorable. Hence, the only grid state with the adjacency matrix $A^{T_\mathrm{B}}$ is $G_H^{T_\mathrm{B}}$, up to addition of $1$-edges, which do not affect $A^{T_\mathrm{B}}$.  
The degree matrix of a (hyper-) graph $H$ is denoted as $D(H)$ and counts in how many hyperedge each vertex is contained in. 
Therefore, $\rho_H^{T_\mathrm{B}} = D(H) + A^{T_\mathrm{B}}$ can only be written as a grid state if $D(H)\ge D({G_H^{T_\mathrm{B}}})$, such that the diagonal operator $\Delta = D(H)-D({G_H^{T_\mathrm{B}}})\ge 0$ accounts for the missing $1$-edges:
\begin{equation}
    \rho_H^{T_\mathrm{B}} = \rho_{G_H^{T_\mathrm{B}}} + \Delta, 
\end{equation}
where $\rho_{G_H^{T_\mathrm{B}}}$ is a grid state of a partially transposed graph. 
\end{proof}
Observe that this proof also implies that $\mathrm{SN}(\rho_H^{T_\mathrm{B}})\le 2$, as an explicit decomposition into states with Schmidt rank $1$ and $2$ is performed.

\section{Schmidt number three PPT state in $5\times 5$}
\label{app:55app}
Like for the crosshatch state in Eq.~\eqref{eq:ch_range}, we explicitly write out the parameterized range of $\rho^{5,5}$, 
\begin{equation}
\Psi^{5,5}= 
    \begin{pmatrix}
        c_{00} & c_{01} & c_{11} & 0 & 0   \\
        c_{10} & c_{11} & c_{12} & 0 & c_{14} \\
        c_{11} & c_{21} & c_{22} & c_{23} & 0 \\
        0 & 0 & c_{32} & c_{22} & c_{12}\\
        0 & c_{41} & 0 & c_{21} & 0 
    \end{pmatrix} \in R(\rho^{5\times 5}), 
\end{equation}
and show that the state $\ket{e_{11}}=\ket{02}+\ket{11}+\ket{20}$ corresponding to the only rank-$3$ edge is orthogonal to $R_2(\rho^{5,5})$. This immediately
implies that this component of $\rho^{5,5}$ cannot be spanned with Schmidt rank $2$ states.
In order to obtain $R_2(\rho^{5,5})$, we now demand that all the $3\times3$ minors of $\Psi^{5,5}$ must vanish. Manifestly, if we can conclude $c_{11}=0$ from this demand, we 
know that $\ket{e_{11}}\perp R_2(\rho^{5,5})$. One of the $3\times3$ vanishing minor conditions translates to the constraint, 
\begin{equation} \label{eq:5by5vanishing_minor}
    c_{11}c_{21}c_{12}=0,
\end{equation}
from which we immediately assume the worst cases, either $c_{21}=0$ or $c_{12}=0$. Both cases are symmetric to another, so we assume that $c_{12}=0$, update the matrix $\Psi^{5,5}$,   
and obtain two more minor equations:
\begin{align}
 c_{11}^2 c_{22} =0 \quad\text{and}\quad
     c_{11}^2 c_{21} =0,
\end{align}
implying that $c_{21},c_{22}=0$, which, proceeding the same way as before, yields the minor
\begin{equation}
    \begin{vmatrix}
        c_{00} & c_{01} & c_{11}\\
        c_{10} & c_{11} & 0\\
        c_{11} & 0 &0
    \end{vmatrix} = c_{11}^3 = 0, 
\end{equation}
implying  $c_{11}=0$ in $R_2(\rho^{5,5})$. Thus, $\rho^{5,5}$ is Schmidt number $3$ PPT bound entangled state.

\section{A constructive method for Schmidt number concentration}\label{app:snfamily}
Let Alice and Bob initially share a bound entangled state $\rho$ with Schmidt number $\mathrm{SN}(\rho)=k$, detectable with the generalized range criterion, and with some additional properties needed for our construction. First, assume that the state $\rho$ can be decomposed as a sum of a sub-normalized state, $\tilde\sigma$, and SR $k$ projector $\ket{\tilde v}\bra{\tilde v}$, 
\begin{equation}
\rho_\mathrm{A_1B_1}=\tilde\sigma_\mathrm{A_1B_1}+\ket{\tilde v}\bra{\tilde v}_\mathrm{A_1B_1}.    
\end{equation}
Additionally, assume that $\ket{\tilde v}\perp R_{k-1}(\rho_\mathrm{A_1B_1})$, with $\bra{\tilde v}\tilde \sigma\ket{\tilde v}=0$ and the existence of a product state $\ket{\alpha\beta}$ in the kernel of $\tilde \sigma$, satisfying $\bra{\alpha\beta}\tilde\sigma\ket{\alpha\beta}=0$ and $\langle\alpha\beta|\tilde v\rangle=t\neq 0$.  For the states satisfying these requirements, we construct a 
map $\Theta$, that takes $\rho_{A_1B_1}$ and tensors it with a $3\times 3$ crosshatch state, and after the local filtering operations returns a higher Schmidt number bound entangled state. Cor convenience we restate the set of edges present in the crosshatch state,  $\{e_1,e_2,e_3,e_4\} = \{\{(1,0),(0,2)\},\{(2,0),(1,2)\},\{(0,0),(2,1)\},\{(0,1),(2,2)\}\}$.

Given a state $\rho_{A_1B_1}$ with SN $k$, the  tensor product state $\rho_{\mathrm{A}_1 \mathrm{B}_1}\otimes \rho^{\mathrm{CH}}_{\mathrm{A}_2 \mathrm{B}_2}$ now has the Schmidt number $2k$, with $\ket{\tilde{v}}_{A_1B_1} \otimes \ket{e_i}_{A_2 B_2}\ \perp\  R_{2k-1} (\rho_{\mathrm{A}_1 \mathrm{B}_1}\otimes \rho^{\mathrm{CH}}_{\mathrm{A}_2 \mathrm{B}_2})$.  
We now define the local filtering operations on Alice's and Bob's sides using the product vector in the kernel of $\tilde\sigma$,
\begin{align}
    A_{\mathrm{A}_1\mathrm{A}_2} &= \ket{\alpha }\bra{\alpha}_{\mathrm{A}_1} \otimes \left(\ket{0}\bra{0}_{\mathrm{A}_2} +\ket{2}\bra{2}_{\mathrm{A}_2}\right)+\mathbb{1}_{\mathrm{A}_1}\otimes\ket{ 1}\bra{1}_{{A}_2},
    \\
    B_{\mathrm{B}_1\mathrm{B}_2} &= \mathbb{1}_{\mathrm{B}_1}\otimes\left(\ket{ 0}\bra{0}_{{B}_2}+\ket{ 1}\bra{1}_{{B}_2}\right) + \ket{\beta }\bra{\beta }_{\mathrm{B}_1}\otimes \ket{2}\bra{2}_{\mathrm{B}_2},
\end{align}
and the following quantities for shortening the subsequent notation:
\begin{alignat}{2}
    \tilde \sigma_{\beta,\mathrm{A}_1} &= \left(\mathbb{1}_{\mathrm{A}_1}\otimes\bra{\beta}_{\mathrm{B}_1}\right)\tilde\sigma_{\mathrm{A}_1\mathrm{B}_1}
    \left(\mathbb{1}_{\mathrm{A}_1}\otimes\ket{\beta}_{\mathrm{B}_1}\right)
    , \quad &
    \ket{\tilde v_{\beta}}_{\mathrm{A}_1} &= \left(\mathbb{1}_{\mathrm{A}_1}\otimes\bra{\beta}_{\mathrm{B}_1}\right) 
    \ket{\tilde v}_{\mathrm{A}_1\mathrm{B}_1},
    \nonumber\\
    \tilde\sigma_{\alpha,\mathrm{B}_1} &= \left(\bra{\alpha}_{\mathrm{A}_1}\otimes\mathbb{1}_{\mathrm{B}_1}\right)\tilde\sigma_{\mathrm{A}_1\mathrm{B}_1}\left(\ket{\alpha}_{\mathrm{A}_1}\otimes\mathbb{1}_{\mathrm{B}_1}\right), &
    \ket{\tilde v_{\alpha}}_{\mathrm{A}_1} &= \left(\bra{\alpha}_{\mathrm{A}_1}\otimes\mathbb{1}_{\mathrm{B}_1}\right)\ket{\tilde v}_{\mathrm{A}_1\mathrm{B}_1}\nonumber\\
    \tilde\tau_{\alpha\beta,\mathrm{A}_1\mathrm{B}_1} &= 
    \left(\bra{\alpha}_{\mathrm{A}_1}\otimes\mathbb{1}_{\mathrm{B}_1}\right)
    \tilde\sigma_{\mathrm{A}_1\mathrm{B}_1}
    \left(\mathbb{1}_{\mathrm{A}_1}\otimes\ket{\beta}_{\mathrm{B}_1}\right).&
    \label{app:eq:notation_family}
\end{alignat}
We put everything together in the following expression depicting the action of $\Theta$,
\begin{align}
\label{app:eq:components_family}
    \Theta(\rho_\mathrm{A_1 B_1})  =&    \left(A_{\mathrm{A}_1\mathrm{A}_2}\otimes B_{\mathrm{B}_1\mathrm{B}_2}\right) \left((\tilde \sigma_{\mathrm{A}_1\mathrm{B}_1}+\ket{\tilde v}\bra{\tilde v})\otimes\rho^{\mathrm{CH}}_{\mathrm{A}_2\mathrm{B}_2}\right)
    \left(A_{\mathrm{A}_1\mathrm{A}_2}\otimes B_{\mathrm{B}_1\mathrm{B}_2}\right)^\dagger=\\
    &{
    \left({\ket{\tilde v}_{\mathrm{A}_1\mathrm{B}_1}\ket{10}_{\mathrm{A}_2\mathrm{B}_2}+t\ket{\alpha\beta}_{\mathrm{A}_1\mathrm{B}_1}\ket{02}_{\mathrm{A}_2\mathrm{B}_2}}\right)\left({\bra{\tilde v}_{\mathrm{A}_1\mathrm{B}_1}\bra{10}_{\mathrm{A}_2\mathrm{B}_2}+t^*\bra{\alpha\beta}_{\mathrm{A}_1\mathrm{B}_1}\ket{02}_{\mathrm{A}_2\mathrm{B}_2}}\right)}
    \nonumber\\
    +&
    \left(\ket{\alpha \tilde v_\alpha}_{\mathrm{A}_1\mathrm{B}_1}\ket{20}+
    \ket{\tilde v_\beta \beta}_{\mathrm{A}_1\mathrm{B}_1}
    \ket{12}_{\mathrm{A}_2\mathrm{B}_2}\right)\left(\bra{\alpha \tilde v_\alpha}_{\mathrm{A}_1\mathrm{B}_1}\bra{20}+
    \bra{\tilde v_\beta \beta}_{\mathrm{A}_1\mathrm{B}_1}
    \bra{12}_{\mathrm{A}_2\mathrm{B}_2}\right)\nonumber
    \\
    +&
    {
    {\ket{\alpha \tilde v_\alpha}
    \bra{\alpha \tilde v_\alpha}_{\mathrm{A}_1\mathrm{B}_1}
    \otimes\ket{e_3}
    \bra{e_3}_{\mathrm{A}_2\mathrm{B}_2}
    }}\nonumber
    \\
    +&\left(\ket{\alpha \tilde v_\alpha}_{\mathrm{A}_1\mathrm{B}_1}\ket{01}_{\mathrm{A}_2\mathrm{B}_2}+t \ket{\alpha\beta}_{\mathrm{A}_1\mathrm{B}_1}
    \ket{12}_{\mathrm{A}_2\mathrm{B}_2}\right)
    \left(\bra{\alpha \tilde v_\alpha}_{\mathrm{A}_1\mathrm{B}_1}\bra{01}_{\mathrm{A}_2\mathrm{B}_2}+t^* \bra{\alpha\beta}_{\mathrm{A}_1\mathrm{B}_1}
    \bra{12}_{\mathrm{A}_2\mathrm{B}_2}\right)
    \nonumber
    \\
    +& \tilde\sigma_{\mathrm{A}_1\mathrm{B}_1} \otimes \ket{1}\bra{1}_{\mathrm{A}_2}\otimes \ket{0}\bra{0}_{\mathrm{B}_2}\nonumber\\
+&\ket{\alpha}\bra{\alpha}_{\mathrm{A}_1}\otimes\tilde\sigma_{\alpha,\mathrm{B}_1} \otimes\ket{2}\bra{2}_{\mathrm{A}_2}\otimes \ket{0}\bra{0}_{\mathrm{B}_2} 
    +\tilde\sigma_{\beta,\mathrm{A}_1}\otimes\ket{\beta}\bra{\beta}_{\mathrm{B}_1} \otimes\ket{1}\bra{1}_{\mathrm{A}_2}\otimes\ket{2}\bra{2}_{\mathrm{B}_2}  \nonumber\\
    +&\left(\ket{\alpha}_{\mathrm{A}_1}\bra{\beta}_{\mathrm{B}_1}\tilde\tau_{\alpha\beta,\mathrm{A}_1\mathrm{B}_1}\otimes\ket{2}\bra{1}_{\mathrm{A}_2}\otimes\ket{0}\bra{2}_{\mathrm{B}_2} + 
\ket{\beta}_{\mathrm{B}_1}\bra{\alpha}_{\mathrm{A}_1}\tilde\tau_{\alpha\beta,\mathrm{A}_1\mathrm{B}_1}^\dagger\otimes\ket{1}\bra{2}_{\mathrm{A}_2}\otimes\ket{2}\bra{0}_{\mathrm{B}_2}\right)
    \nonumber\\
    +&{
    {\ket{\alpha}\bra{\alpha}_{\mathrm{A}_1}\otimes\tilde\sigma_{\alpha,\mathrm{B}_1} \otimes\ket{e_3}\bra{e_3}_{\mathrm{A}_2\mathrm{B}_2}}} 
    +\ket{\alpha}\bra{\alpha}_{\mathrm{A}_1}\otimes\tilde\sigma_{\alpha,\mathrm{B}_1}\otimes \ket{0}\bra{0}_{\mathrm{A}_2}\otimes \ket{1}\bra{1}_{\mathrm{B}_2}.
\end{align}

One can work out individual terms to notice that the resulting state on subsystems $\mathrm{A_1 A_2}$ and $\mathrm{B_1B_2}$ can be written as a convex combination of a subnormalized mixed state $\tilde{\eta}$ with SN at most $k$, and a Schmidt rank $(k+1)$ sub-normalized projector, 
\begin{equation}
     \Theta(\rho_{\mathrm{A}_1\mathrm{B}_1})= \tilde{\eta} +\ket{\tilde{w}}\bra{\tilde{w}}, 
\end{equation}
where $\ket{\tilde w}=A_{\mathrm{A}_1\mathrm{A}_2}\otimes B_{\mathrm{B}_1\mathrm{B}_2}\ket{\tilde{v}}_{A_1B_1}\otimes \ket{e_1}=\ket{\tilde v}_{\mathrm{A}_1\mathrm{B}_1}\ket{10}_{\mathrm{A}_2\mathrm{B}_2}+t\ket{\alpha\beta}_{\mathrm{A}_1\mathrm{B}_1}\ket{02}_{\mathrm{A}_2\mathrm{B}_2}$ evidently has Schmidt rank $(k+1)$. So, $\Theta$  preserves the structure of the state. We have to, however, show that other assumed properties of $\rho$ are preserved too. To this end, we  observe that
\begin{equation}       \left(\mathbb{1}_{\mathrm{A}_1\mathrm{B}_1}\otimes\bra{10}_{\mathrm{A}_2\mathrm{B}_2}\right)
\Theta(\rho_\mathrm{A_1 B_1})    \left(\mathbb{1}_{\mathrm{A}_1\mathrm{B}_1}\otimes\ket{10}_{\mathrm{A}_2\mathrm{B}_2}\right)=\Tilde{\sigma}_{\mathrm{A}_1\mathrm{B}_1} + \ket{\Tilde{v}}\bra{\Tilde{v}}_{\mathrm{A}_1\mathrm{B}_1}={\rho}_{\mathrm{A}_1\mathrm{B}_1},
\end{equation}
which we term \textit{nesting property} of $\Theta$, since it shows that ${\rho}_{\mathrm{A}_1\mathrm{B}_1}$ is 'nested' inside $\Theta(\rho_{\mathrm{A}_1\mathrm{B}_1})$ and can be recovered with a local projection.
Therefore,  $\ket{\alpha1\beta0}_{\mathrm{A}_1\mathrm{B}_1\mathrm{A}_2\mathrm{B}_2}$ can always be chosen as product state fulfilling the same requirements for $\Theta(\rho_{\mathrm{A}_1\mathrm{B}_1})$, as $\ket{\alpha\beta}_{\mathrm{A}_1\mathrm{B}_1}$does for $\rho_{\mathrm{A}_1\mathrm{B}_1}$

Next, since $\rho^\mathrm{CH}\ket{11}=0$, from the structure of local filtering, one can see that there exists a block of kernel elements in the image of $\Theta$: 
\begin{equation}
\Theta(\rho_\mathrm{A_1 B_1})\left(\mathbb{1}_{\mathrm{A}_1\mathrm{B}_1}\otimes\ket{1}\bra{1}_{\mathrm{A}_2}\otimes\ket{1}\bra{1}_{\mathrm{B}_2}\right)=0.
\end{equation}
Assuming $\ket{\psi}_{\mathrm{A_1}\mathrm{B_1}\mathrm{A_2}\mathrm{B_2}}\in R_k(\Theta(\rho_\mathrm{A_1 B_1}))$, in order to show an increased SN, we aim to show $\ket{\psi}_{\mathrm{A_1}\mathrm{B_1}\mathrm{A_2}\mathrm{B_2}}\perp \ket{\tilde v}_{\mathrm{A_1}\mathrm{B_1}\mathrm{A_2}\mathrm{B_2}}$, 
by considering size $(k+1)$ minors given by a selection of $k$ distinct $\{i_a\}_{1\ge a \ge k} = \{i_1,\dots i_{k}\}$, $\{j_b\}_{1\ge b \ge k} = \{j_1,\dots j_{k-1}\}$ and an arbitrary normalized state $\ket{x}_{\mathrm{B}_1}$:
\begin{align}
&\left|
    \begin{matrix}
        \left(\mean{i_a j_b10|{\psi}}_{\mathrm{A_1}\mathrm{B_1}\mathrm{A_2}\mathrm{B_2}}\right)_{ab} & \left(\mean{i_a x11|{\psi}}_{\mathrm{A_1}\mathrm{B_1}\mathrm{A_2}\mathrm{B_2}}\right)_{a}\\
        \left(\mean{\alpha j_b 20|{\psi}}_{\mathrm{A_1}\mathrm{B_1}\mathrm{A_2}\mathrm{B_2}}\right)_{b} & 
        \mean{\alpha x 21|{\psi}}_{\mathrm{A_1}\mathrm{B_1}\mathrm{A_1}\mathrm{B_1}}
    \end{matrix}\right| = \left|
    \begin{matrix}
        \left(\mean{i_a j_b10|{\psi}}_{\mathrm{A_1}\mathrm{B_1}\mathrm{A_2}\mathrm{B_2}}\right)_{ab} & 0_{(k-1)\times 1}\\
        \left(\mean{\alpha j_b 20|{\psi}}_{\mathrm{A_1}\mathrm{B_1}\mathrm{A_2}\mathrm{B_2}}\right)_{b} & 
        \mean{\alpha x 21|{\psi}}_{\mathrm{A_1}\mathrm{B_1}\mathrm{A_2}\mathrm{B_2}}
    \end{matrix}\right|=\nonumber\\ &=
    \mean{\alpha x 21|{\psi}}_{\mathrm{A_1}\mathrm{B_1}\mathrm{A_2}\mathrm{B_2}}\left| \left(\mean{i_a j_b10|{\psi}}_{\mathrm{A_1}\mathrm{B_1}\mathrm{A_2}\mathrm{B_2}}\right)_{ab} \right|=0.
\end{align}
In order to simultaneously fulfill this family of equations, either $\left| \left(\mean{i_a j_b10|{\psi}}_{\mathrm{A_1}\mathrm{B_1}\mathrm{A_2}\mathrm{B_2}}\right)_{ab} \right|=0$ must hold, regardless of the choice of $i_a$, $j_b$, or $\mean{\alpha x 21|{\psi}}_{\mathrm{A_1}\mathrm{B_1}\mathrm{A_2}\mathrm{B_2}}$ must hold, regardless of the choice of $\ket{x}_{\mathrm{B}_1}$. Assuming the first case, and using the nesting property of $\Theta(\rho_{\mathrm{A}_1\mathrm{B}_1})$, it follows that
\begin{align}
\left(\mathbb{1}_{A_1 B_1}\otimes\bra{10}_{A_2 B_2}\right)&\ket{\psi}_{\mathrm{A_1}\mathrm{B_1}\mathrm{A_2}\mathrm{B_2}}\in R(\rho_{\mathrm{A_1}\mathrm{B_1}}),\\
\left(\mathbb{1}_{A_1 B_1}\otimes\bra{10}_{A_2 B_2}\right)&\ket{\psi}_{\mathrm{A_1}\mathrm{B_1}\mathrm{A_2}\mathrm{B_2}}\perp \ket{\Tilde{v}}_{\mathrm{A_1}\mathrm{B_1}}.
\end{align}
The second equation implies $\ket{\psi}_{\mathrm{A_1}\mathrm{B_1}\mathrm{A_2}\mathrm{B_2}}\perp \ket{\Tilde{w}}_{\mathrm{A_1}\mathrm{B_1}\mathrm{A_2}\mathrm{B_2}}$, as evident from the structure of $\Theta(\rho_{\mathrm{A}_1\mathrm{B}_1})$. For the case $\mean{\alpha x 21|{\psi}}_{\mathrm{A_1}\mathrm{B_1}\mathrm{A_2}\mathrm{B_2}}=0$, it follows that 
\begin{align}  \ket{{\psi}}_{\mathrm{A_1}\mathrm{B_1}\mathrm{A_2}\mathrm{B_2}}&\perp
    R(\ket{\alpha}\bra{\alpha}_{\mathrm{A}_1}
    \otimes\left(\tilde\sigma_{\alpha,\mathrm{B}_1}+\ket{\tilde v_\alpha}\bra{\tilde v_\alpha}_{\mathrm{B}_1}\right)\otimes
    \ket{e_3}\bra{e_3}_{\mathrm{A}_2\mathrm{B}_2})\\
    \implies \ket{{\psi}}_{\mathrm{A_1}\mathrm{B_1}\mathrm{A_2}\mathrm{B_2}}&\perp \ket{\alpha 00}_{\mathrm{A_1}\mathrm{A_2}\mathrm{B_2}}.
\end{align}
This causes another family of minors (using the same convention for $i_a$, $j_b$) to simplify:
\begin{align}
&\left|
    \begin{matrix}
        \left(\mean{\alpha j_b 00|{\psi}}_{\mathrm{A_1}\mathrm{B_1}\mathrm{A_2}\mathrm{B_2}}\right)_{b} & \mean{\alpha\beta 02|{\psi}}_{\mathrm{A_1}\mathrm{B_1}\mathrm{A_2}\mathrm{B_2}}\\
        \left(\mean{i_a j_b 10|{\psi}}_{\mathrm{A_1}\mathrm{B_1}\mathrm{A_2}\mathrm{B_2}}\right)_{ab} & 
        \left(\mean{i_a \beta 12|{\psi}}_{\mathrm{A_1}\mathrm{B_1}\mathrm{A_2}\mathrm{B_2}}\right)_{a}
    \end{matrix}\right| = \left|
    \begin{matrix}
        0_{1\times(k-1)} & \mean{\alpha\beta 02|{\psi}}_{\mathrm{A_1}\mathrm{B_1}\mathrm{A_2}\mathrm{B_2}}\\
        \left(\mean{i_a j_b 10|{\psi}}_{\mathrm{A_1}\mathrm{B_1}\mathrm{A_2}\mathrm{B_2}}\right)_{ab} & 
        \left(\mean{i_a \beta 12|{\psi}}_{\mathrm{A_1}\mathrm{B_1}\mathrm{A_1}\mathrm{B_1}}\right)_{a}
    \end{matrix}\right|=\nonumber\\ &=
    -\mean{\alpha\beta 02|{\psi}}_{\mathrm{A_1}\mathrm{B_1}\mathrm{A_2}\mathrm{B_2}}
    \left|\left( \mean{i_a j_b10|{\psi}}_{\mathrm{A_1}\mathrm{B_1}\mathrm{A_2}\mathrm{B_2}}\right)_{ab} \right|=0.
\end{align}
If we assume that $\ket{\alpha\beta 02}_{\mathrm{A_1}\mathrm{B_1}\mathrm{A_2}\mathrm{B_2}}\perp\ket{{\psi}}_{\mathrm{A_1}\mathrm{B_1}\mathrm{A_2}\mathrm{B_2}}$, it implies that $\ket{\Tilde{w}}_{\mathrm{A_1}\mathrm{B_1}\mathrm{A_2}\mathrm{B_2}}\perp\ket{{\psi}}_{\mathrm{A_1}\mathrm{B_1}\mathrm{A_2}\mathrm{B_2}}$, since $\ket{\Tilde{w}}\bra{\Tilde{w}}_{\mathrm{A_1}\mathrm{B_1}\mathrm{A_2}\mathrm{B_2}}$ is the only component of $\Theta(\rho_{\mathrm{A}_1\mathrm{B}_1})$ overlapping with $\ket{\alpha\beta 02}_{\mathrm{A_1}\mathrm{B_1}\mathrm{A_2}\mathrm{B_2}}$. For the other case (all minors $\left|  \left(\mean{i_a j_b10|{\psi}}_{\mathrm{A_1}\mathrm{B_1}\mathrm{A_2}\mathrm{B_2}}\right)_{ab} \right|$ vanish), it was already established that $\ket{\Tilde{w}}_{\mathrm{A_1}\mathrm{B_1}\mathrm{A_2}\mathrm{B_2}}\perp\ket{{\psi}}_{\mathrm{A_1}\mathrm{B_1}\mathrm{A_2}\mathrm{B_2}}$ holds. We have thus proven that
\begin{equation}
    \ket{\Tilde{w}}_{\mathrm{A_1}\mathrm{B_1}\mathrm{A_2}\mathrm{B_2}}\perp R_k(\Theta(\rho_\mathrm{A_1 B_1})),
\end{equation}
and, therefore, that $\mathrm{SN}(\Theta(\rho))= k+1$.

\section{Schmidt number $(d+1)/2$ grid states in $d\times (d+1)(d+3)/8  $}\label{app:distil_Grid}
The map $\Theta$ introduced in Appendix~\ref{app:snfamily}, Eq.~\eqref{app:eq:components_family} increases the first local dimension by $d_\mathrm{A}\mapsto d_\mathrm{A} + 2$ with each application, 
while the increase of the second local dimension appears to be $d_\mathrm{B}\mapsto 2d_\mathrm{B} +1$. 
If it is possible to choose $\ket{\alpha}$, such that the matrix rank $\mathrm{rank}(\Tilde{\sigma}_\alpha + \ket{\Tilde{v}_\alpha}\bra{\Tilde{v}_\alpha}) = r_\mathrm{B}< d_\mathrm{B}$, an increase of 
$d_\mathrm{B}\mapsto d_\mathrm{B}+r_\mathrm{B} +1$ can be achieved by Bob choosing a projection that removes components $\ket{y}_\mathrm{B}$ satisfying $\Theta(\rho_\mathrm{AB})\mathbb{1}_\mathrm{A}\otimes\ket{y}_\mathrm{B}=0$. Such components correspond to empty columns in the grid state formalism. To this end, Bob also projects on $R((\Tilde{\sigma}_{\alpha,\mathrm{B}_1} + \ket{\Tilde{v}_\alpha}\bra{\Tilde{v}_\alpha}_{\mathrm{B}_1})\otimes\ket{1}\bra{1}_{\mathrm{B}_2} + \mathbb{1}_\mathrm{B_1}\otimes(\ket{0}\bra{0}_{\mathrm{B}_2}+\ket{2}\bra{2}_{\mathrm{B}_2}))$. 
The quantities $\Tilde{\sigma}_{\alpha,\mathrm{B}_1}$, $\ket{\Tilde{v}_\alpha}_{\mathrm{B}_1}$ are the same as those defined in Eq.~\eqref{app:eq:notation_family} in Appendix~\ref{app:snfamily}.
This is possible, since for any $\ket{y}_\mathrm{B_1}$ in the kernel of $\Tilde{\sigma}_\alpha+\ket{\Tilde{v}_\alpha}\bra{\Tilde{v}_\alpha})$, it holds that $\Theta(\rho_\mathrm{A_1 B_1})\mathbb{1}_{\mathrm{A}_1\mathrm{A}_2}\otimes\ket{y1}_{\mathrm{B}_1\mathrm{B}_2}=0$. 
Afterward, the procedure can be iterated by choosing $\ket{\alpha}_{\mathrm{A}_1} \mapsto \ket{\alpha 1}_{\mathrm{A}_1\mathrm{A}_2}$, $\ket{\beta}_{\mathrm{B}_1} \mapsto \ket{\beta 0}_{\mathrm{B}_1\mathrm{B}_2}$, where $\mathrm{rk}\left(\bra{\alpha 1}_{\mathrm{A}_1\mathrm{A}_2}\Theta(\rho_\mathrm{A_1 B_1})\ket{\alpha 1}_{\mathrm{A}_1\mathrm{A}_2}\right) = r_\mathrm{B}+1$, since $\ket{x1}_{\mathrm{B_1}\mathrm{B_2}}\perp R(\bra{\alpha 1}_{\mathrm{A}_1\mathrm{A}_2}\Theta(\rho_\mathrm{A_1 B_1})\ket{\alpha 1}_{\mathrm{A}_1\mathrm{A}_2})$, for any $\ket{x}_{\mathrm{B}_1}$. 
As shown in Appendix \ref{app:snfamily}, this choice also guarantees that all assumptions are met in every step.
\begin{align}
    \rho^{(1)}_\mathrm{AB} &= \rho_{\mathrm{A_1 B_1}}\\
    \rho^{(m+1)}_\mathrm{AB} &= \Theta(\rho^{(m)}_\mathrm{A_1 B_1}) = \Theta^{m}(\rho^{(1)}_\mathrm{A_1 B_1}),
\end{align}
Using this scheme, with $m$ counting applications of $\Theta$, we can observe that 
\begin{alignat}{2}
    d_\mathrm{B}(\rho^{(m+1)}_\mathrm{AB}) &= d_\mathrm{B}(\rho^{(m)}_\mathrm{AB}) + r_\mathrm{B}(\rho^{(m)}_\mathrm{AB}) +1, & \quad r_\mathrm{B}(\rho^{(m+1)}_\mathrm{AB}) &= r_\mathrm{B}(\rho^{(m)}_\mathrm{AB}) +1 \implies \\
    d_\mathrm{B}(\rho^{(m+1)}_\mathrm{AB}) &= d_\mathrm{B}(\rho^{(1)}_\mathrm{AB}) + m r_\mathrm{B}(\rho^{(1)}_\mathrm{AB}) +\frac{m(m+1)}{2}, & \quad r_\mathrm{B}(\rho^{(m+1)}_\mathrm{AB}) &=  r_\mathrm{B}(\rho^{(1)}_\mathrm{AB}) +m
\end{alignat}
yielding quadratic scaling, even assuming $r_\mathrm{B}(\rho^{(1)}_\mathrm{AB})=d_\mathrm{B}(\rho^{(1)}_\mathrm{AB})$. 
When we insert the crosshatch state, $\rho^{(1)}_\mathrm{AB}=\rho^{\mathrm{CH}}_\mathrm{AB}$, 
and choose the initial $\ket{\alpha\beta}_{\mathrm{AB}} = \ket{10}_{\mathrm{AB}}$ and $\ket{v}_{\mathrm{AB}} = \ket{e_1}_{\mathrm{AB}}$, all assumptions required for distillation are fulfilled. The local dimensions thus obtained are $d_\mathrm{A} = (3+2m)$, $d_\mathrm{B} = (m+2)(m+3)/2$, and the Schmidt number is $(2+m)$.
With the previously stated tensor product and filtering rules, the family of states can be represented as grid states, shown in figure~\ref{fig:snfamily}.
The Schmidt number scaling ($(d+1)/2$) achieved with this augmentation map method is not only 
an improvement over the previously best known $d/2$ scaling, 
but the method itself may be viewed as an instance of bound entanglement assisted distillation~\cite{ishizaka2004effect}, applied to Schmidt number concentration in the PPT set.

\section{Schmidt number three PPT state in $4\times 12$}\label{app:4x12}
The scheme described in the previous section can be slightly modified to obtain a Schmidt number $3$ bound entangled state $\rho^{4,12}$ in local dimensions $4\times 12$. We start with the tensor product between the crosshatch state and the bound entangled state $\rho^{\mathrm{Hor}}$ in $2\times4$,  known as Horodecki state~\cite{horodecki1997separability}, which has the highest known relative Schmidt number $\mathrm{SN}(\rho^{\mathrm{Hor}})/d_{\min}=1$. 
\begin{align}
\begin{split}
    \rho^{\mathrm{Hor}}=\frac{1}{10}(\ket{00}\bra{00}+(\ket{01}+\ket{10})(\bra{01}+\bra{10})+(\ket{03}+\ket{12})(\bra{03}+\bra{12})\\
    (\ket{02}+\ket{11})(\bra{02}+\bra{11})+(\sqrt{2}\ket{13}+\ket{10})(\sqrt{2}\bra{13}+\bra{10})) .
\end{split}
\end{align}
\begin{figure}[H]
    \centering
    \includegraphics[width=0.25\textwidth]{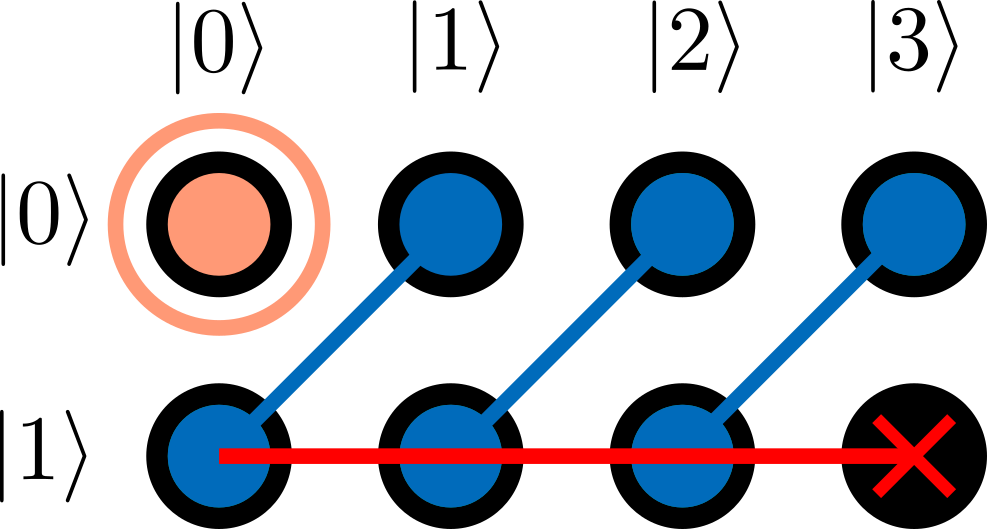}
    \caption{Grid-state-like representation of $\rho^{\mathrm{Hor}}$, with the weighted edge indicated with an X-shape at the end with the higher weight. All other $1$- and $2$-edges can be understood the same way as in a grid state.}
    \label{app:fig:HorState}
\end{figure}

As it is evident,   $\rho^{\mathrm{Hor}}$ is not a grid state, as not all states in the convex mixture correspond to equal superpositions of grid vertices. Instead, we have one \textit{weighted edge}, $\sqrt{2}\ket{13}+\ket{10}$. It is, however, still convenient to depict such a state on a grid (especially after the distillation protocol). See Fig.~\ref{app:fig:HorState} for our representation of  $\rho^{\mathrm{Hor}}$. Moreover, note that we have chosen a particular representative for the family of states given in \cite{horodecki1997separability}, as this allows representing $\rho^{\mathrm{Hor}}$ similar to a grid state; other states in the family are equally suitable for constructing high Schmidt number states in the same fashion.

In order to obtain $\rho^{4,12}$, we perform a different local filtering map acting only on Alice's side, on the tensor product with the $3\times 3$ crosshatch state, $\rho^{\mathrm{Hor}}\otimes\rho^{\mathrm{CH}}$: 
\begin{equation}
    \rho^{4,12}_{{A,B}} =  \left(\mathbb{1}_{\mathrm{B}}\otimes\Pi_{\mathrm{A}}\right)\rho^{\mathrm{Hor}}_{\mathrm{A}_1\mathrm{B}_1}\otimes\rho^{\mathrm{CH}}_{\mathrm{A}_2\mathrm{B}_2}  
    \left(\mathbb{1}_{\mathrm{B}\otimes\Pi_{\mathrm{A}}}\right)^\dagger, 
\end{equation}
where $ \Pi_{\mathrm{A}} := \mathbb{1}_{\mathrm{A}_1\mathrm{A}_2}-\ket{1}\bra{1}_{\mathrm{A}_1}\otimes\left(\ket{0}\bra{0}+\ket{2}\bra{2}\right)_{\mathrm{A}_2} $. To ease the notation, we relabel Alice's basis as follows $\{\ket{00}_{\mathrm{A_1}\mathrm{A_2}},\ket{01}_{\mathrm{A_1}\mathrm{A_2}},\ket{11}_{\mathrm{A_1}\mathrm{A_2}},\ket{02}_{\mathrm{A_1}\mathrm{A_2}}\}\quad \mapsto\quad\{\ket{0}_{\mathrm{A}},\ket{1}_{\mathrm{A}},\ket{2}_{\mathrm{A}},\ket{3}_{\mathrm{A}}\}$ and Bob's basis as $\ket{ij}_{\mathrm{B}_1\mathrm{B}_2}\mapsto \ket{i+4j}_{\mathrm{B}}$. As a result, we obtain a PPT state in $4\times 12$, which has a grid state like representation, with an additional weighted edge, as was in the original state $\rho^{\mathrm{Hor}}$. The state after the mapping is given by the equal mixture of the subsequent (unnormalized) states:

\begin{alignat}{2}
\ket{\Tilde{s}} &= \sqrt{2}\ket{2,0}+\ket{2,3},\quad &  
\ket{\Tilde{s}'} & = \sqrt{2}\ket{2,8}+\ket{2,11},\nonumber\\
\{\ket{h_i} &= \ket{0,9+i} + \ket{1,i}+\ket{2,1+i}\}_{0\le i \le 2},\quad & 
\ket{h_3} &= \ket{0,1}+\ket{3,8},\nonumber\\ 
\{\ket{r_i} &= \ket{1,9+i} + \ket{2,8+i}+\ket{3,1+i}\}_{0\le i \le 2},\quad&
\ket{r_3} &= \ket{0,3}+\ket{1,8},\nonumber \\
\{\ket{q_i} &= \ket{0,4+i}+\ket{3,i}\}_{0\le i \le 7}. & &
\end{alignat}

\begin{figure}[H]
    \centering
    \includegraphics[width=0.7\textwidth]{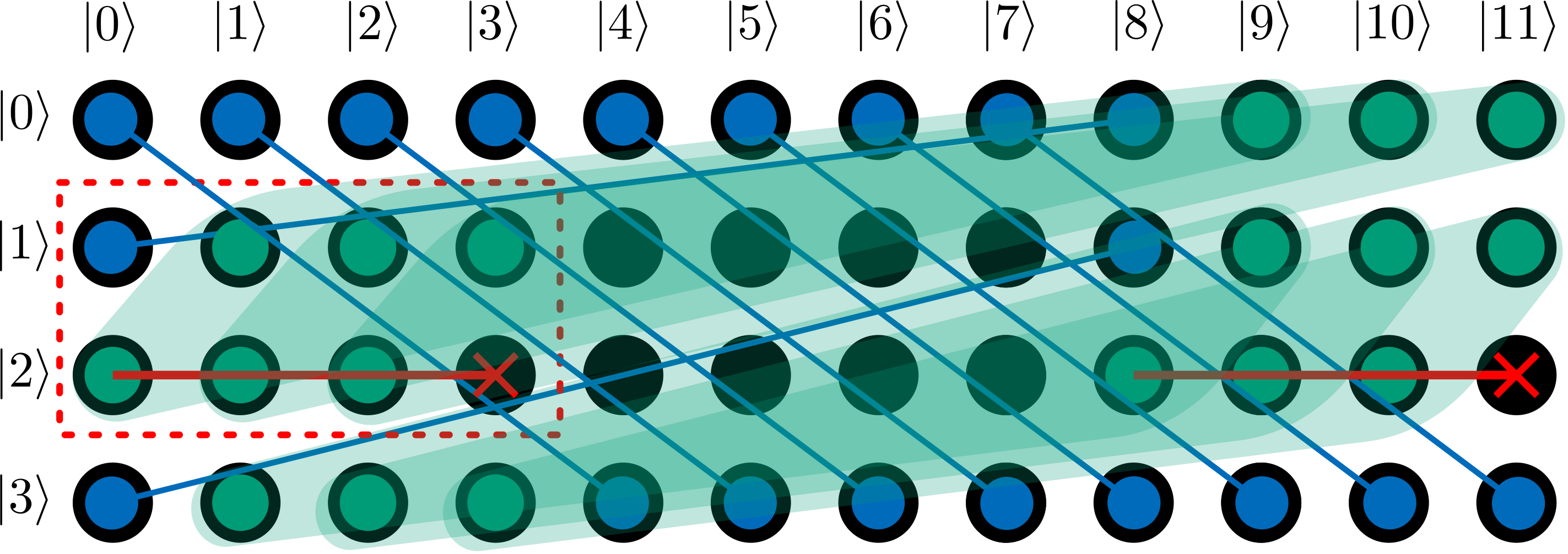}
    \caption{Representation of $\rho^{4,12}$ in the manner of a grid state. It is obtained by equal mixture of the components
    $\{\ket{h_i} = \ket{0,9+i} + \ket{1,i}+\ket{2,1+i}\}_{0\le i \le 2}$, 
$\ket{h_3} = \ket{0,1}+\ket{3,8}$, $\ket{\Tilde{s}} = \sqrt{2}\ket{2,0}+\ket{2,3}$, $\{\ket{q_i}= \ket{0,4+i}+\ket{3,i}\}_{0\le i \le 7}$, 
$\{\ket{r_i} = \ket{1,9+i} + \ket{2,8+i}+\ket{3,1+i}\}_{0\le i \le 2}$, 
$\ket{r_3} = \ket{0,3}+\ket{1,8}$, $\ket{\Tilde{s}'} = \sqrt{2}\ket{2,8}+\ket{2,11}$.
    The states $\ket{\Tilde{s}}$ and   $\ket{\Tilde{s}'}$, which do not correspond to valid hyperedges are represented with red edges. Note the isolated vertices, and the Horodecki state block indicated by the square.}
    \label{app:fig:4x12sn3}
\end{figure}

See Fig.~\ref{app:fig:4x12sn3} for the grid-like state and its caption for the list of edges and two weighted edges. Note that this state has a similar nesting property as the previous family:
\begin{align}
    &P_{\mathrm{AB}}  = 
    \left(\sum_{1\le i \le2} \ket{i}\bra{i}_{\mathrm{A}}
    \right)\otimes
    \left(\sum_{j\le 3} \ket{j}\bra{j}_{\mathrm{B}}
    \right),\qquad\qquad P_{\mathrm{AB}}\rho^{4,12}_{\mathrm{AB}}P_{\mathrm{AB}}  \propto \rho^{\mathrm{Hor}}_{\mathrm{AB}}.
\end{align}

Thus, if we suppose that $\rho^{4,12}$  can be decomposed into Schmidt rank $2$ states, at least one state $\ket{\psi} \in R_2(\rho)$ in the decomposition must have a Schmidt rank $2$ projection $P\ket{\psi}$. 
Suppose that for all states $\ket{\psi_i}$ in the decomposition $\rho^{4,12}_{\mathrm{AB}} = \sum_i p_i \ket{\psi_i}\bra{\psi_i}$, the projections $P_{\mathrm{AB}}\ket{\psi_i}$ are product states, then it would follow that 
\begin{equation}
    P_{\mathrm{AB}}\rho^{4,12}_{\mathrm{AB}}P_{\mathrm{AB}} = \sum_i p_i P_{\mathrm{AB}}\ket{\psi_i}\bra{\psi_i}P_{\mathrm{AB}} \propto \rho^{\mathrm{Hor}}_{\mathrm{AB}},
\end{equation}
wrongly implying that $\rho^{\mathrm{Hor}}_{\mathrm{AB}}$ is separable. This proves the previous statement by contradiction.

This Schmidt rank $2$ projection can be associated to a size $2\times2$ block, with $i_0,i_1\le 3$, with nonzero determinant: 
\begin{equation}
\left|
    \begin{matrix}
        \mean{1i_0|\psi}_{\mathrm{AB}}& \mean{1i_1|\psi}_{\mathrm{AB}}\\
        \mean{2i_0|\psi}_{\mathrm{AB}} & \mean{2i_1|\psi}_{\mathrm{AB}}
    \end{matrix}\right|\neq 0.
    \label{app:eq:nonzero_det}
\end{equation}

Due to the central block of kernel elements $\{ \ket{ij}\}_{1\le i \le2,\: 4\le j\le 7}$,
there are size $3$ minors for all $4\le i_3\le 7$, of the form:
\begin{alignat}{2}
    \left|\begin{matrix}
        \mean{0i_0|\psi}_{\mathrm{AB}}& \mean{0i_1|\psi}_{\mathrm{AB}} & \mean{0i_3|\psi}_{\mathrm{AB}}\\
        \mean{1i_0|\psi}_{\mathrm{AB}}& \mean{1i_1|\psi}_{\mathrm{AB}} & 0\\
        \mean{2i_0|\psi}_{\mathrm{AB}} & \mean{2i_1|\psi}_{\mathrm{AB}} & 0
    \end{matrix}\right| = 
    \mean{0i_3|\psi}_{\mathrm{AB}}
        \left|\begin{matrix}
        \mean{1i_0|\psi}_{\mathrm{AB}}& \mean{1i_1|\psi}_{\mathrm{AB}}\\
        \mean{2i_0|\psi}_{\mathrm{AB}} & \mean{2i_1|\psi}_{\mathrm{AB}}
    \end{matrix}\right|=0
    &\implies \langle{0i_3|\psi}\rangle_{\mathrm{AB}}&=0
    \\
    \left|\begin{matrix}
        \mean{1i_0|\psi}_{\mathrm{AB}}& \mean{1i_1|\psi}_{\mathrm{AB}} & 0\\
        \mean{2i_0|\psi}_{\mathrm{AB}} & \mean{2i_1|\psi}_{\mathrm{AB}} & 0\\
           \mean{3i_0|\psi}_{\mathrm{AB}}& \mean{3i_1|\psi}_{\mathrm{AB}} & \mean{3i_3|\psi}_{\mathrm{AB}}\\
    \end{matrix}\right| = 
    \mean{3i_3|\psi}_{\mathrm{AB}}
        \left|\begin{matrix}
        \mean{1i_0|\psi}_{\mathrm{AB}}& \mean{1i_1|\psi}_{\mathrm{AB}}\\
        \mean{2i_0|\psi}_{\mathrm{AB}} & \mean{2i_1|\psi}_{\mathrm{AB}}
    \end{matrix}\right|=0
    &\implies \langle{3i_3|\psi}_{\mathrm{AB}}\rangle&=0.
\end{alignat}
This implies that $\ket{\psi}$ must be orthogonal to all $\ket{q_i}$, and hence to all $\ket{0 i_4}$ with $i_4\le 3$. Thus, the following minors involving $8\le i_5$ also factorize:
\begin{equation}
\left|\begin{matrix}
        0 & 0 & \mean{0i_5|\psi}_{\mathrm{AB}}\\
        \mean{2i_0|\psi}_{\mathrm{AB}} & \mean{2i_1|\psi}_{\mathrm{AB}} & \mean{1i_5|\psi}_{\mathrm{AB}}\\
           \mean{3i_0|\psi}_{\mathrm{AB}}& \mean{3i_1|\psi}_{\mathrm{AB}} & \mean{2i_5|\psi}_{\mathrm{AB}}\\
    \end{matrix}\right| =
    \mean{0i_5|\psi}_{\mathrm{AB}}
    \left|\begin{matrix}
        \mean{2i_0|\psi}_{\mathrm{AB}} & \mean{2i_1|\psi}_{\mathrm{AB}} \\
           \mean{3i_0|\psi}_{\mathrm{AB}}& \mean{3i_1|\psi}_{\mathrm{AB}} \\
    \end{matrix}\right| =0 \implies \langle{0i_5|\psi}\rangle_{\mathrm{AB}}=0.
\end{equation}
Since $\ket{\psi}\perp \ket{h_i}$, it follows  that $P_{\mathrm{AB}}\ket{\psi}\propto \ket{\Tilde{s}}$, contradicting Eq.~\eqref{app:eq:nonzero_det}.
Thus, for every $\ket{\psi}\in R_2(\rho^{4,12}_{\mathrm{AB}})$, $P_{\mathrm{AB}}\ket{\psi}$ is a product state, and it is not possible to find a decomposition of $\rho^{4,12}_{\mathrm{AB}}$ into pure states in 2-restricted range, as that would imply a decomposition of the Horodecki state into product states.
We can therefore conclude that $ \rho^{4,12}$ has Schmidt number $3$.

\section{Edge States of High Schmidt Number}
\label{app:snedgestates}
\begin{figure}
    \centering
    \includegraphics[width=0.45\textwidth]{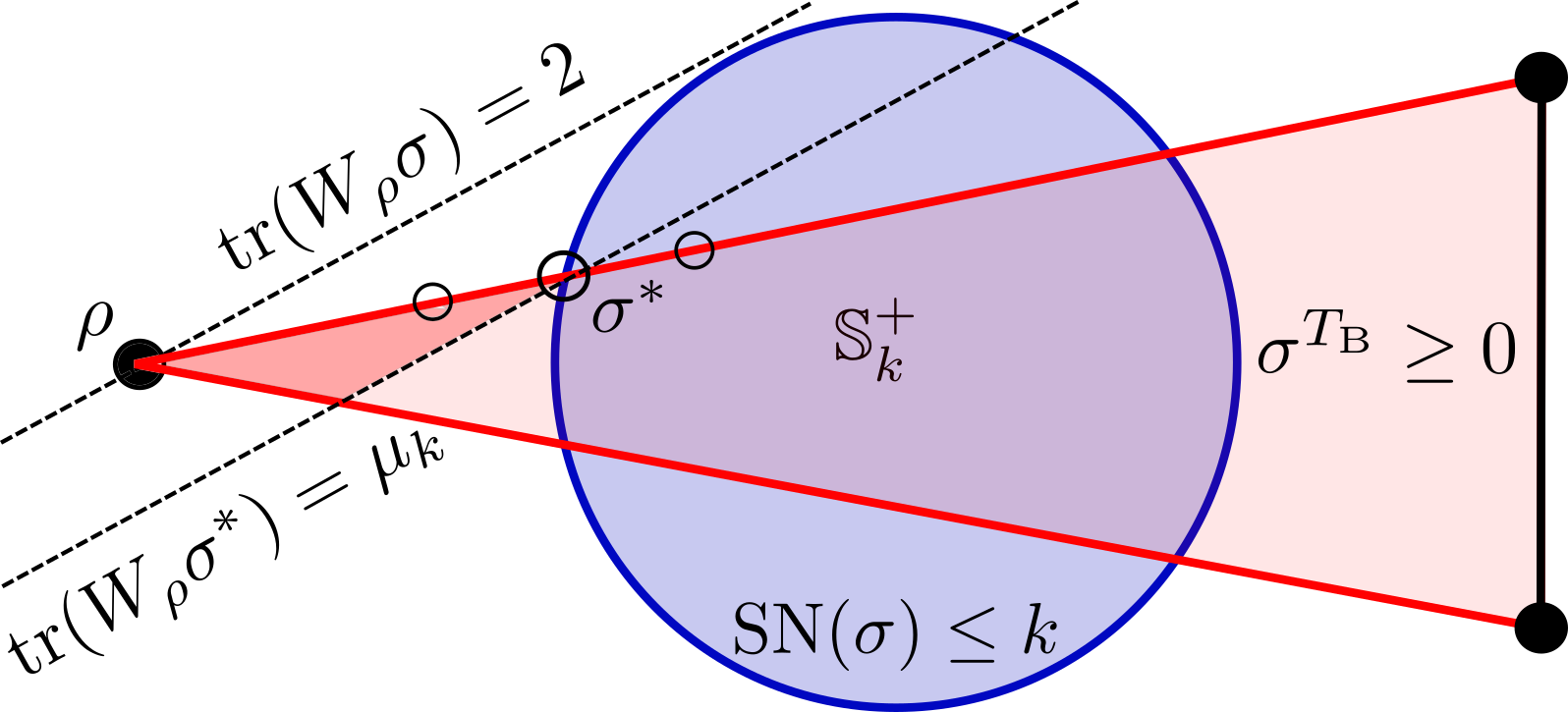}
    \caption{
    Schematic relationship between Schmidt number $k$ set (shown as shaded circle), PPT set (shown as triangle), and the witness $W_\rho$ (shown as dotted lines) discussed in this section. A $\rho$ that is extremal to the PPT set is shown, and the optimum $\sigma^*=\arg\max_{\sigma\in\mathbb{S}_k^+}\tr(W_\rho\sigma)$ is a nontrivial extremal state of $\mathbb{S}_k^+$, with a possible decomposition indicated with small circles. Edge states to $\mathbb{S}_k^+$ are indicated in black.}
    \label{fig:pptset}
\end{figure}

The witness construction in the main text is not only applicable to extremal states, but to any state $\rho$ which stops to be PPT as soon as a (subnormalized) Schmidt number $k$ PPT state $\Tilde\sigma_k$ is subtracted from it. Such states are referred to as \textit{edge states} to the Schmidt number $k$ PPT set, and are a straightforward generalization of edge states as introduced in \cite{chruscinski2008construct,lewenstein2001characterization}. Naturally, any $\Tilde\sigma$ that can be subtracted from $\rho$, while preserving the PPT property, has $R(\Tilde\sigma)\subseteq R(\rho)$ and $R(\Tilde\sigma^{T_\mathrm{B}})\subseteq R(\rho^{T_\mathrm{B}})$. Hence, $W_\rho = P+ Q^{T_\mathrm{B}}$ also detects $\sigma=\Tilde{\sigma}/\tr(\Tilde{\sigma})$ with $\tr(W_\rho\sigma)=2$, and since $\rho$ is an edge state, all such $\sigma$ have a Schmidt number $>k$. Hence, there must be a bound $\mu_k$ for the PPT Schmidt number $k$ set $\mathbb{S}_k^+$, such that \begin{equation}
    \mu_k:=\max\limits_{\sigma_k \in \mathbb{S}_k^+} \tr{W_\rho\sigma_k}<2.
\end{equation}
Thus it is also possible to construct Schmidt number witnesses out of edge states to PPT Schmidt number $k$ sets. See Fig.~\ref{fig:pptset} for the relationship between quantities associated to $W_\rho$ and edge states.
We further note that the bound $\mu_k$ can always be saturated by a $\sigma_k$ that is an extremal state to $\mathbb{S}_k^+$. As the only PPT states that are also extremal states of the  Schmidt number $k$ set (i.e., pure states of Schmidt rank not higher than $k$) are product states again, all other extremal states arise from the intersection of a face of the Schmidt number $k$ set with a face of the PPT set. Nontrivial extremal states that are neither extremal to the Schmidt number $k$ set nor to the PPT set seem highly desirable to construct, as every such state $\rho$ allows a convex decomposition into PPT states $\rho=t\sigma_1+(1-t)\sigma_2$, one of which has a Schmidt number of at least $k+1$. Hence, for example, the problem of finding a Schmidt number $3$ PPT state in local dimensions $3\times 4$ can be reduced to finding a Schmidt number $2$ state with the aforementioned properties, without a need to construct any $W_\rho$.

\section{Testing for Extremality}
\label{app:extremetest}
For any PPT $\rho$, its extremality can be numerically determined by intersecting the corresponding hyperplanes defined by $R(\rho)$ and $R(\rho^{T_\mathrm{B}})$ and confirming that they intersect in a unique ray~\cite{leinaas2007extreme}. To this end, we define the following projections acting on the set of hermitian matrices: 
\begin{align}
    \mathbf{P}(\sigma) =  P\sigma P & & \mathbf{Q}(\sigma) =  (Q\sigma^{T_\mathrm{B}} Q)^{T_\mathrm{B}},
\end{align}
where $P$,$Q$ are again projections on $R(\rho)$, $R(\rho^T_{\mathrm{B}})$, hence $\mathbf{P},\mathbf{Q}$ project on the set of hermitian operators supported on the two ranges. 
For an extremal state, this intersection is unique, hence there only is a single joint eigenvector $\rho$, such that $\mathbf{P}(\rho)=\rho$ and $\mathbf{Q}(\rho)=\rho$ \cite{leinaas2007extreme}. Equivalently, making use of positive semidefiniteness of projections, $\mathbf{P}+\mathbf{Q}$ has a nondegenerate eigenvalue of $2$, which is confirmed for all examples presented in table \ref{tab:mutable}.

\section{Determining $\mu_1$ on some examples}
\label{app:numrange}
To upper-bound the separable numerical range of an operator $W$, it is sufficient to solve $\max_{\ket{xy}}\bra{xy}W\ket{xy}$ for product states \cite{lewenstein2000optimization}. We apply the DPS hierarchy to obtain a lower bound, and determine upper bounds and likely optima using local optimization with a high number ($10^7$) of starting points.

For local optimization, the following seesaw algorithm \cite{ling2010biquadratic} was used.
The initial point $\ket{x^{(0)}}$ is chosen randomly, and then updated in the following manner, where $\arg\max$ always is the eigenvector associated to the largest eigenvalue:
\begin{align}
    \ket{y^{(i)}} &= \arg\max\limits_{\ket{y}_\mathrm{B}} \bra{y}_\mathrm{B} \left(\bra{x^{(i)}}_\mathrm{A}\otimes\mathbb{1}_\mathrm{B}\right)W \left(\ket{x^{(i)}}\otimes\mathbb{1}_\mathrm{B}\right)
    \ket{y}_\mathrm{B}
    \\
    \ket{x^{(i+1)}} &= \arg\max\limits_{\ket{x}_\mathrm{A}} \bra{x}_\mathrm{A}\left(\mathbb{1}_\mathrm{A}\otimes\bra{y^{(i)}}_\mathrm{B}\right)W \left(\mathbb{1}_\mathrm{A}\otimes\ket{y^{(i)}}_\mathrm{B}\right)\bra{x}_\mathrm{A}
\end{align}
This method always converges to a local optimum, as every step is guaranteed to decrease $\bra{xy}W\ket{xy}$. 
The local optimality condition may be written as a collection of $d_\mathrm{A}+d_\mathrm{B}$ degree three polynomial equations with $d_\mathrm{A}+d_\mathrm{B}+1$ complex variables in the form of the generalized eigenvalue problem \cite{sperling2009representation}:
\begin{align}
    \left(\mathbb{1}_\mathrm{A}\otimes\bra{y}_\mathrm{B}\right) W \ket{xy} = \lambda \ket{x}\\
    \left(\bra{x}_\mathrm{A}\otimes\mathbb{1}_\mathrm{B}\right) W \ket{xy} = \lambda \ket{y},
\end{align}
allowing to bound the number of local optima as $3^{d_\mathrm{A}+d_\mathrm{B}+1}$, using Bézout's theorem. Note that local optimum must be understood as a connected component of solutions, not a single isolated point. In analogy to the ordinary eigenvalue problem, every connected component has $\lambda$ constant.
As $\bra{xy}W\ket{xy}$ is infinitely differentiable everywhere, and local optima of different values are separated from another, it is in principle possible to determine the global optimum with a sufficient number of starting points. 
The DPS hierarchy \cite{doherty2004complete} was implemented as the SDP
\begin{align}
    &\min_\rho \qquad\tr(\rho_{\mathrm{A}\mathrm{B}_1\dots \mathrm{B}_n} W_{\mathrm{A}\mathrm{B}_1})\\
    &\mathrm{s.t.}\qquad \quad  P^\vee_{\mathrm{B}_1\dots \mathrm{B}_n} \rho_{\mathrm{A}\mathrm{B}_1\dots \mathrm{B}_n} P^\vee_{\mathrm{B}_1\dots \mathrm{B}_n} = \rho, \\
    &\qquad \qquad\ \tr\rho=1,\ \rho\ge 0,\ \rho^{T_\mathrm{A}}\ge 0,
\end{align}
where $P^\vee$ is the projector on the symmetric subspace. As is customary, the constraint involving the projector is implemented as parametrization, in order to reduce the number of optimization variables. In all cases, the extension was performed on the smaller of the two subsystems to further reduce the number of required variables. The optimization was performed using a first order method, the Splitting Cone Solver \cite{scs}, in order to lower space requirements and allow a larger number of extensions.

\begin{table}[H]
    \centering
    \begin{tabular}{c|ccccc}
         & $\rho^{\mathrm{CH}}$ 
         &$\rho^{5,5}$ & $\rho^{(2)}$ &$\rho^{(3)}$ & $\rho^{4,12}$\\
        $\mu_\mathrm{U}-\mu_\mathrm{L}$ & $9.42\times10^{-5}$ &  $8.94\times 10^{-4}$ & 
        $3.75\times10^{-8}$ & 
        $3.63\times 10^{-2}$ &
        $4.63\times10^{-10}$ \\
        $\mu_\mathrm{L}$ & $\approx 1.8659$ &  $1.9$ & $\frac{5+\sqrt{5}}{4}$ &$\frac{5+\sqrt{5}}{4}$ & $\frac{5+2\sqrt{2}}{4}$\\
        level & $8$ & $3$ & $2$ & $3$ & $6$\\
        max. \# of optima & $2187$ &   $177147$ & $387420489$x &$531441$ & $129140163$\\
        
    \end{tabular}
    \caption{Table containing upper and lower bounds of $\mu_1$ for various states, alongside the level $n$ used in the DPS hierarchy to determine the upper bound. The table contains an upper bound on the number of local maxima determined by Bézout's theorem.
    For every example, the numerical lower bound suggested an analytical solution that was confirmed to be a stationary point of the optimization problem, and the associated $\lambda$ was given in closed form.
    For $\rho^{\mathrm{CH}}$, the analytical estimate obtained from local optimization is $\frac{1}{10}(9+\sqrt{8}+\sqrt{41+2^{5/2}})$.}
    \label{tab:mutableapp}
\end{table}
\end{appendix}

\twocolumngrid
\bibliography{bibl.bib}
\end{document}